\documentclass[aps,prl,twocolumn,superscriptaddress,numerical,amsmath,amssymb,floatfix]{revtex4-2}
\usepackage[utf8]{inputenc}
\usepackage[english]{babel}
\usepackage[T1]{fontenc}
\usepackage{MnSymbol}
\usepackage{amsmath,dsfont,amsfonts,bbm,pifont}
\usepackage[colorlinks,
citecolor=black,
linkcolor=black,
urlcolor=black]{hyperref}
\usepackage{tikz}
\usepackage{physics}
\usepackage{braket}
\usepackage{orcidlink}
\usepackage{multirow}
\usepackage[export]{adjustbox}

\usepackage{overpic}
\usepackage{xcolor}
\usepackage{graphicx}\usepackage{dcolumn}\usepackage{bm}\usepackage{verbatim}

\renewcommand{\Tr}{\mathrm{Tr}}

\begin{document}

\title{
From Effective Temperature to Non-Boltzmann State Selection in Driven-Dissipative Quantum Criticality}

\author{T. Jorge}
\affiliation{CeFEMA-LaPMET, Departamento de Física, Instituto Superior Técnico, Universidade de Lisboa, Avenida Rovisco Pais, 1049-001 Lisboa, Portugal}

\author{J. Paaske}
\affiliation{Niels Bohr Institute, University of Copenhagen, 2100 Copenhagen, Denmark}

\author{P. Ribeiro}
\affiliation{CeFEMA-LaPMET, Departamento de Física, Instituto Superior Técnico, Universidade de Lisboa, Avenida Rovisco Pais, 1049-001 Lisboa, Portugal}

\date{\today}

\begin{abstract}

Whether a nonequilibrium quantum-critical steady state can be described by an effective temperature remains an open question. We address it in a voltage-biased electronic Lipkin--Meshkov--Glick model, a minimal driven-dissipative model of a collective spin coupled to metallic leads.  A controlled large-$N$ treatment reveals an
overdamped open quantum-critical regime distinct from the closed model. The surrounding quantum critical fan, organized by temperature and voltage, is explored and fluctuations are found to be governed by an effective temperature $T_{\rm eff}$. At strong bias, the transition becomes first-order through a tricritical point.  Remarkably, the same $T_{\rm eff}$ governs the strongly driven regime, but now varies across the entire order-parameter landscape. The steady state is therefore selected by a non-Boltzmann rule, shifting the first-order transition away from the equal-depth point of a deterministic potential. Driven criticality thus remains organized by an effective temperature while revealing non-thermal state selection.

\end{abstract}

\maketitle

Voltage bias, temperature gradients, and other drives impose nonequilibrium electronic distributions, producing steady states whose correlations and susceptibilities generally lack an equilibrium counterpart. Thus, the universal behavior organized by quantum critical points (QCPs) in equilibrium correlated matter~\cite{SachdevQPT, Vojta_2003, Coleman_2005_Criticality, FrerotRoscilde2019, Gegenwart2008_HeavyFermionQCP, Keimer2015_HighTcQPT, Dutta2015_QPT_SpinModels} need not survive under nonequilibrium driving. For slow, soft degrees of freedom, fluctuation-dissipation ratios (FDRs) can nevertheless be reduced to an effective equilibrium description with an effective temperature $T_{\rm eff}$ set by bath temperature and drive strength~\cite{Cugliandolo1997_EffectiveTemperatures, Mitra_2006, Mitra_2008, Ribeiro2013, Ribeiro2015}. This reduction is expected to be most accurate near phase transitions, where critical slowing down makes the order parameter evolve on the longest time scale.

Whereas criticality in Markovian driven-dissipative systems has seen important progress, notably through renormalization-group  analyses and universality classifications~\cite{Sieberer2013, Marcuzzi2014_RydbergNoneq, Sieberer_2016, Maghrebi2016Jan, Fink_2017, Rota2019Mar, SiebererDiehl2025,tauber2014critical}, many-body criticality with non-Markovian environments is more diverse and remains understood mainly through examples~\cite{Dalidovich2004Jul, Green2005Dec, Feldman2005Oct, Mitra_2006, Mitra_2008, Sela2009Jan, Kirchner2009Nov, DeVega2017_NonMarkovReview, Schaller2014_OpenQS, chakraborty_sensarma_2018, Walldorf2019Sep, Lundgren2020Sep, Zhang2021Feb, Kawamura2025Dec}. This regime is especially relevant for electronic transport, where reservoirs both dissipate and impose nonthermal occupations. Moreover, since mean-field descriptions often fail in driven open systems~\cite{Maghrebi2016Jan, Lundgren2020Sep, Walldorf2019Sep, Okugawa2026Jan}, finding critical models where the roles of drive and dissipation become transparent remains an important task~\cite{Chung2009May, Sela2009Jan, Kirchner2009Nov, Zhang2021Feb}. A central question is therefore when an effective-equilibrium description can be defined near a QCP, what its predictive power is, {and what it leaves undetermined}.

\begin{figure}[!t]
\centering
\begin{tabular}{@{}c@{}}
\includegraphics[width=0.963\linewidth]{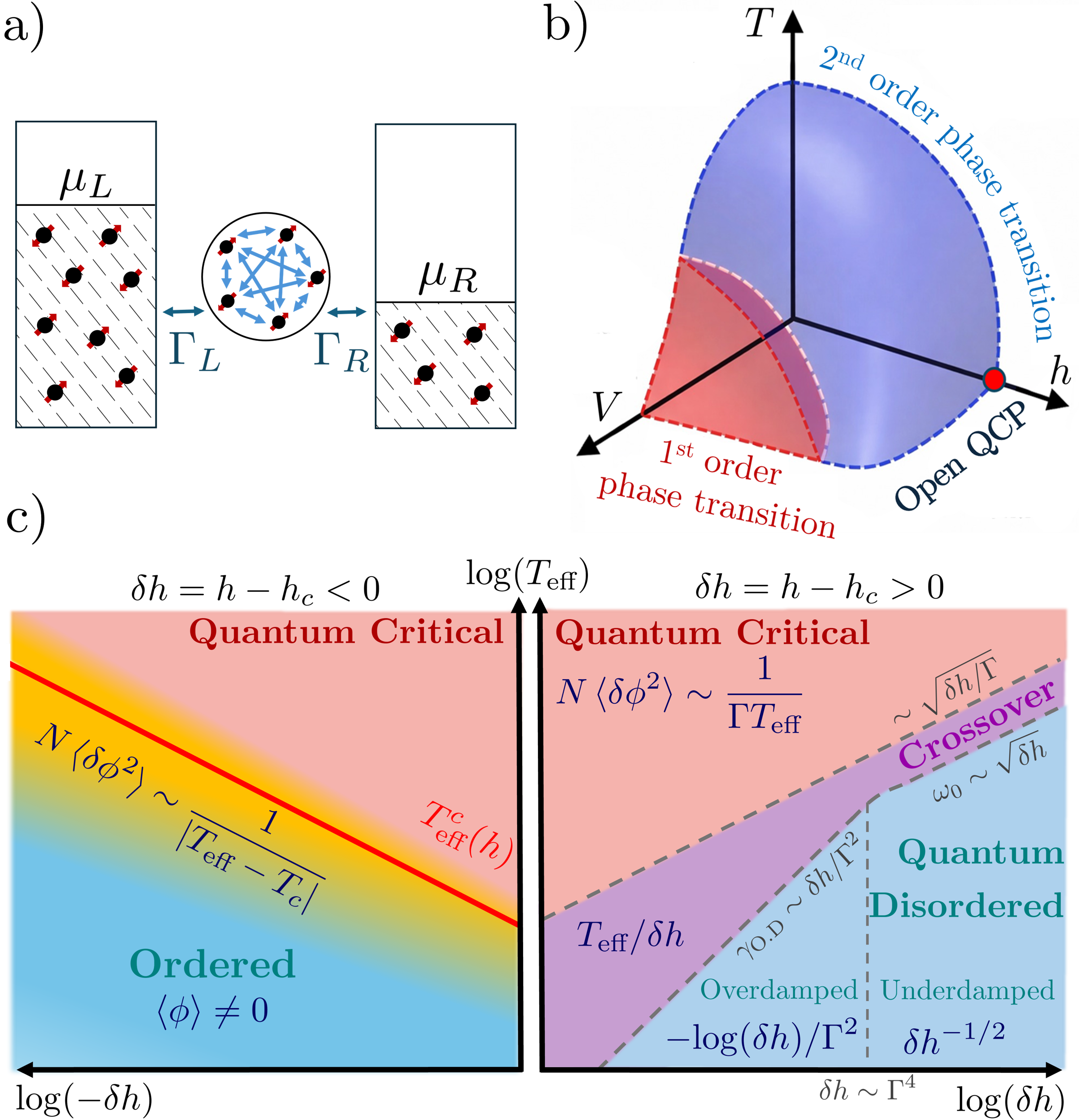}
\end{tabular}
\caption{\textbf{Model and nonequilibrium phase diagram.}
\textbf{(a)} Schematic of the electronic LMG model: a magnetic quantum dot
hosting a collective spin, tunnel-coupled to two metallic leads held at
chemical potentials $\mu_L =-\mu_R=V/2$ with tunnelling rates $\Gamma_{L}=\Gamma_{R}=\Gamma/2$.
\textbf{(b)} Sketch of the phase diagram for $\phi=\langle\hat S_x\rangle/N$ in
$(h,T,V)$ space at small $\Gamma$: the blue region is ordered ($\phi=0$
unstable), the red region bistable ($\phi=0$ and $\phi\neq0$ both stable or
metastable), and the red dot at $V=T=0$ marks the open QCP, with distinct scaling laws from the closed LMG. \textbf{(c)} Sketch of the quantum-critical fan in the
$(\delta h,\,T_{\mathrm{eff}})$ plane [See Eq.~(\ref{eq:teff})],
covering both the ordered ($\delta h<0$) and disordered ($\delta h>0$) sides, with the scaling of $N\langle\delta\phi^2\rangle$ indicated in each region. The
quantum-critical regime is common to both sides; at low $T_{\mathrm{eff}}$ the
quantum-disordered regime hosts an overdamped/underdamped crossover, also common to the deep-ordered regime. Fluctuations diverge on the critical line $T_{\mathrm{eff}}^c(h)$ \vspace{-5 mm}}
\label{fig:main}
\end{figure}

In this Letter, we answer this question for an instructive model featuring a magnetic quantum dot whose collective spin realizes a Lipkin--Meshkov--Glick (LMG) interaction~\cite{LIPKIN1965} and is coupled to voltage-biased metallic leads [Fig. ~\ref{fig:main} (a)], with the nonequilibrium phase diagram sketched in Fig.~\ref{fig:main}(b). The large-$N$ limit provides a controlled nonequilibrium Green's-function treatment in which hybridization strength, temperature, and bias enter on equal footing. Three elements substantiate our findings, each highlighting an important feature of driven-dissipative criticality. First, opening the system is itself a relevant critical perturbation: even at $T=V=0$, the hybridization $\Gamma$ modifies the scaling of the closed collective-spin transition, producing an overdamped open QCP controlled by $\delta h/\Gamma^4$~\cite{Yin2014, Vacari_2019_PhysRevA.100.052108, Rossini_2021}. {Second, the effective-temperature regime extends beyond $T\gg V$: even at $T=0$ the low-frequency Gaussian mode obeys an effective thermal FDR, in the sense that critical fluctuations are organized by a single $T_{\rm eff}(T,V)$ (itself a property of the saddle point about which they are computed) [Fig.~\ref{fig:main}(c)].} {Third, stronger voltage turns the transition first order [Fig.~\ref{fig:main}(b)]. The same effective temperature reappears there, now sampled across the order-parameter landscape rather than at one saddle point; out of equilibrium it varies across the landscape and informs a non-Boltzmann state selection.}

\paragraph*{Electronic Lipkin--Meshkov--Glick (eLMG) Model.---} We study an electronic realization of the Lipkin--Meshkov--Glick model~\cite{LIPKIN1965}. It describes electrons in $N$ orbitals coupled through an infinite-range anisotropic exchange interaction and a perpendicular magnetic field $h$. The collective spin is
$\vec{\hat S}=(1/2)\sum_{i,ss'}\hat{d}_{i,s}^{\dagger}\vec{\sigma}_{ss'}\hat{d}_{i,s'}$, and the closed dot Hamiltonian is
\begin{equation}\label{eq:LMG}
\hat{H}_\mathrm{LMG} = -h\,\hat{S}_z - \frac{J}{N}\,\hat{S}_x^2.
\end{equation}
Originally introduced in nuclear physics~\cite{LIPKIN1965}, such collective-spin models now appear  in settings ranging from cavity-QED and cold atoms~\cite{CavityQED_LMG, LMG_BEC} to single-molecule magnets~\cite{Garanin_1998_SingleMoleculeMagnets, Bartolome2014Molecular, Bode2012Mar, GanzhornWernsdorfer2021}, metallic nanoparticles, and chaotic quantum dots. In electronic nanostructures, transport experiments on ferromagnetic cobalt nanoparticles have been described using collective anisotropic spin exchange~\cite{Gueron1999, CanaliMacDonald2000, Kleff2001}. Related collective spin-exchange physics, including anisotropic Ising couplings~\cite{Lyubshin_2014} and mesoscopic Stoner instabilities~\cite{Andreev1998Oct, Brouwer1999Nov, Waintal2003Dec, Gorokhov2004Apr, Basko2009Feb, Sharafutdinov2014Nov}, has been studied within the universal quantum-dot Hamiltonian ~\cite{Alhassid_2000_QD_StatisticalT, UniversalQDHamiltonian2}. Here, we neglect charging effects, assumed screened in the open regime,
and take degenerate dot levels, focusing on the spin-exchange-driven transition.
This electronic LMG system is tunnel-coupled to two metallic reservoirs held at temperature $T$ and biased by different chemical potentials. The resulting eLMG model is an open, non-Markovian, driven-dissipative system, for which the large-$N$ limit keeps fluctuations controlled.

The full Hamiltonian is
$\hat{H}_\text{eLMG}=\hat{H}_\text{LMG}+\hat{H}_\text{T}+\hat{H}_\text{RES}$, with non-interacting reservoirs
$\hat{H}_\text{RES}=\sum_{l,i,k,s}(\varepsilon_{k}-\mu_{l})\hat{c}^{\dagger}_{liks}\hat{c}_{liks}$
and tunneling
$\hat{H}_{\text{T}} =\sum_{liks}(t_l\hat{c}^{\dagger}_{liks}\hat{d}_{is} + t_l^\ast\hat{d}^{\dagger}_{is}\hat{c}_{liks})$.
Here $\hat{c}^{\dagger}_{liks}$ creates an electron in channel $i$ of lead $l=L,R$, and each dot level couples to an independent, equivalent lead channel with amplitude $t_l$, giving tunneling rates
$\Gamma_l=2\pi\nu_F|t_l|^2$, where $\nu_F$ is the density of states per channel and spin near the Fermi level. The leads have a common temperature and symmetric chemical potentials $\mu_L=-\mu_R=V/2$; for $\Gamma_R=\Gamma_L=\Gamma/2$, the dot remains at half filling,
$\sum_{is}\braket{\hat{d}^\dag_{is}\hat{d}_{is}}/N=1$.

Crucially, the coupling to the leads breaks conservation of the dot's total spin $[\hat{H}_{\text{T}},\hat{S}^2]\neq0$. This is the microscopic origin of the altered critical behavior we report, and it distinguishes the open transport setting from a collective spin subject to decay~\cite{FerreiraRibeiro2019_LMG_Dissipation, KopylovSchaller2019}. This model resembles that of Ref.~\cite{Burmistrov2020May}, but the spin anisotropy in $\hat{H}_{\mathrm{LMG}}$ simplifies the dynamics and makes the two-lead driven problem tractable. Since the eLMG model replaces local exchange by a collective long-range interaction, its phase transition is directly related to influential studies of voltage-biased itinerant ferromagnets~\cite{Feldman2005Oct, Mitra_2006, Mitra_2008}.

The transverse field $h$ competes with the anisotropic exchange $J$, which favors ordering along $x$. The corresponding order parameter is $\phi=\langle \hat{S}_x\rangle/N$. In the closed model, the $\mathbb{Z}_2$ symmetry $\hat{S}_{x,y}\to -\hat{S}_{x,y}$ is broken for $h<J$, yielding two degenerate states with $\phi=\pm|\phi|$, while for $h>J$ the ground state is paramagnetic. The transition is mean-field-like, becoming exact as $N\to\infty$ due to the all-to-all exchange interaction~\cite{BotetJullien1983}; fluctuations are therefore organized as $1/N$ corrections. Throughout, we work in units where $\hbar=k_{B}=J=1$, expressing all energies in units of the exchange coupling.
To assess this phase transition under nonequilibrium conditions, the model is  recast into a Keldysh path-integral~\cite{Kamenev2023Jan}. We arrive at an effective bosonic description by decoupling the exchange interaction with a real bosonic Hubbard--Stratonovich field $\phi$ that captures the collective $x$-polarization of the dot. After integrating out the individual electrons of both the dot and the leads, the resulting action takes the form
\begin{equation}
S = N\!\int\! \mathrm{d}t\,\Big[-2\bm{\phi}^{T}\hat\gamma^{q}\bm{\phi} - i\,\Tr\ln\!\big(i(\hat{G}_0^{-1}-\hat{\Sigma}+\hat{\bm{\phi}}\sigma_x)\big)\Big],
\label{eqn: bosonic action}
\end{equation}
where $\bm\phi=(\phi^{\mathrm{cl}},\phi^{q})^{T}$ carries classical and quantum components, $\hat\gamma^q=\sigma_1$ in Keldysh space, and $\hat{\bm{\phi}}=\phi^{\mathrm{cl}}+\phi^{q}\hat\gamma^q$. Here $\Tr$ runs over time, spin, and Keldysh indices; below, $\Tr_s$ denotes spin trace only. For brevity, we shall henceforth suppress the superscript on $\phi^{\mathrm{cl}}$. The classical field is the dynamical order parameter, with $\langle\phi\rangle=\langle\hat{S}_x\rangle/N$, while $\braket{\phi^q}=0$.
All dependence on temperature and bias voltage enters via the tunneling self-energies given by
\begin{equation}
\Sigma^{R/A}=\mp i\frac{\Gamma}{2},
\quad
\Sigma^{K}(\omega)=-i\!\sum_{l}\frac{\Gamma}{2}\tanh\!\Big[\tfrac{\beta}{2}(\omega-\mu_l)\Big],
\label{eqn: self energies}
\end{equation}
which determine the dressed QD-electron Green functions via the Dyson equation as $(G^{R/A})^{-1}=G_{0}^{-1}-\Sigma^{R/A}$ and $G^K=G^{R}\Sigma^{K}G^{A}$. 
In the thermodynamic limit ($N\to\infty$), the steady-state value of $\bm\phi$ is obtained from the self-consistent saddle-point equation with $\phi^q_0=0$\footnote{The self-consistency comes from the saddle-point dependence in $G_0^R = (\omega+h\sigma_z/2+\phi\sigma_x+i0)^{-1}$.}:
\begin{equation}
\phi=-\frac{i}{4}\int\frac{\mathrm{d}\omega}{2\pi}\Tr_s[\sigma_xG^K(\omega)].
\label{eq:saddle_general}
\end{equation}
We classify transitions through non-analyticities of
$\phi$~\cite{Minganti2018_SpectralLiouvillian,
Kessler2012_SpinDissipation, Debecker2025}: jumps mark first-order transitions, divergent susceptibilities mark continuous (second-order) ones.

\paragraph*{Dissipative Quantum Critical Point.---}

\begin{figure}[!t]
  \centering
  \includegraphics[width=0.88\linewidth]{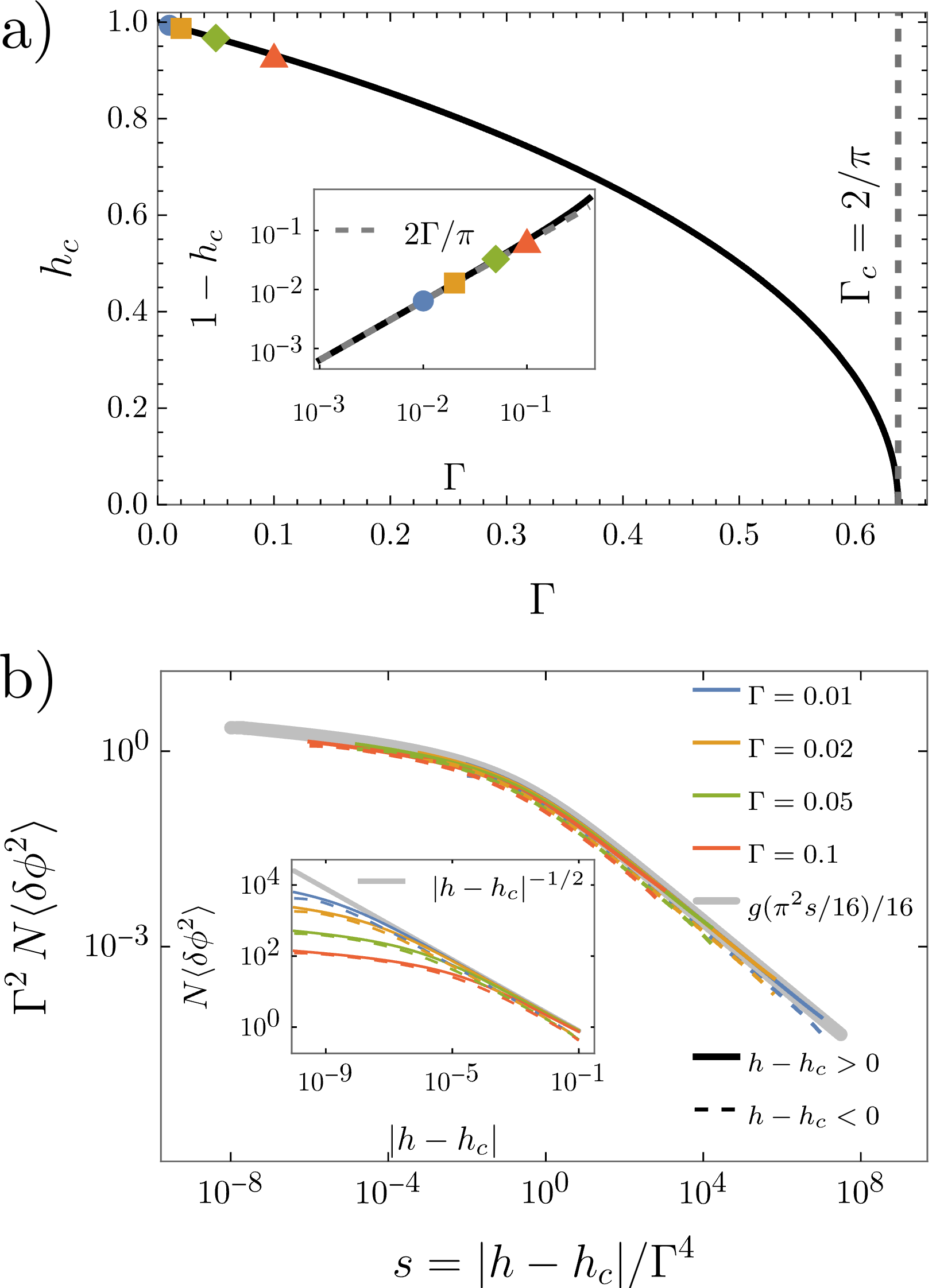}
  \caption{\textbf{Open QCP.} \textbf{(a)} 
  Critical field $h_c(\Gamma)$ separating the ordered ($\phi\neq0$, below) and
  paramagnetic ($\phi=0$, above) phases at $T=V=0$. Colored markers at
  $\Gamma=\{0.01,0.02,0.05,0.1\}$ label the curves used in (b).
  \textbf{(b)} Scaling collapse of the steady-state order-parameter fluctuations: rescaled
  $\Gamma^2 N\langle\delta\phi^2\rangle$ versus $s=|h-h_c|/\Gamma^4$ for the marked values of $\Gamma$,
  on both the ordered ($h-h_c<0$, solid) and disordered ($h-h_c>0$, dashed) sides. Curves
  collapse onto the closed-form scaling function $g$ of Eq.~\eqref{eq:F_function}, with prefactors derived from the values of $a$, $\lambda$ and $c$ in Eq.~\eqref{eq:lowfreq_expansion}. Inset: raw fluctuations $N\langle\delta\phi^2\rangle$ versus
  $|h-h_c|$ before rescaling, showing the logarithmic divergence of the open QCP; the
  closed-model power law $|h-h_c|^{-1/2}$ is shown for comparison.}
  \label{fig:QCP}
\end{figure}

We first focus on the open quantum critical point at $T=V=0$. Coupling to the leads shifts the (second-order) transition: the critical field $h_c$ is suppressed with increasing $\Gamma$ ($h_c\simeq 1-2\Gamma/\pi$ at small $\Gamma$), and the ordered phase is destroyed altogether beyond $\Gamma_c=2/\pi$ [Fig.~\ref{fig:QCP}(a)]. This phenomenon is analogous to the tunneling-induced suppression of the mesoscopic Stoner instability in open quantum dots \cite{Burmistrov2020May}. We henceforth measure the detuning from this dissipation-renormalized critical field, $\delta h\equiv h-h_c(\Gamma)$.
 
Expanding the action~\eqref{eqn: bosonic action} to second order in the fluctuation $\delta\phi=\phi-\langle\phi\rangle$ gives a collective propagator $D$, derived in the End Matter. At quadratic level, criticality is reached when the retarded propagator develops a zero-frequency pole, and the steady-state order-parameter fluctuations follow from the Keldysh component:
\begin{equation}
N\langle\delta\phi^2\rangle=\frac{1}{2}\int\!\frac{\mathrm{d}\omega}{2\pi}\,
\frac{-i\,\Pi^K(\omega)}{\big|\Pi^R(\omega)-2\big|^2}   ,
\label{eq:fluct_integral}
\end{equation}
where criticality corresponds to $\Pi^R(0)=2$, i.e.  $\Pi^R(0)=2/J$ with the
exchange coupling restored. At $T=V=0$, $\Pi^K(\omega)\sim|\omega|$, so that the $\omega\to0$ and $h\to h_c$ limits do not commute, and the fluctuations diverge only logarithmically, $ N\langle\delta\phi^2\rangle \sim -\log|\delta h|$, 
instead of the $|\delta h|^{-1/2}$ power law of the closed LMG model \cite{Dusuel2004,Dusuel_2005}, visible in the inset of Fig.~\ref{fig:QCP}(b).
 
This change signals a distinct dissipative critical scaling regime. In the closed model, the $\delta h^{-1/2}$ spin fluctuation divergence translates, through a scaling-function analysis~\cite{Dusuel2004, Dusuel_2005}, into the well-known finite-size scaling exponent $\langle\delta\phi^2\rangle\sim N^{-2/3}$ at criticality. At the open QCP, the vanishing susceptibility exponent instead yields a leading $\langle\delta\phi^2\rangle\sim N^{-1}$, which we establish in the End Matter through a tree-level renormalization-group analysis that recovers $\langle\delta\phi^2\rangle\sim N^{-2/3}$ in the $\Gamma\to0$ limit.

For weak dissipation, $\Gamma\ll 1$, its influence on the critical behavior may be inferred explicitly from the low-frequency behavior of the inverse fluctuation propagator:
\begin{equation}
2-\Pi^{R/A}(\omega)=a{\mp} i  \lambda\omega-c\omega^2+\mathcal O(\omega^3),
\label{eq:lowfreq_expansion}
\end{equation}
where $a\simeq2|\delta h|/h$, $\lambda\simeq 8\Gamma^2/(\pi h^4)$, and $c\simeq2/h^3$, {up to leading order in $\delta h$ and $\Gamma$}. Since $ac/\lambda^2\sim|\delta h|/\Gamma^4$, Eq.~\eqref{eq:fluct_integral} exhibits a single-variable scaling form,
\begin{equation}
N\langle\delta\phi^2\rangle\;\sim\;\frac{1}{\Gamma^2}\,g\!\left(\frac{|\delta h|}{\Gamma^4}\right),
\label{eq:scaling_form}
\end{equation}
with $g(s)\sim s^{-1/2}$  ($s\gg1$) to $g(s)\sim-\log s$ ($s\ll1$), in agreement with Fig.~\ref{fig:QCP}(b). Dynamically, it marks the change from an underdamped mode with $\omega_0\sim\sqrt{|\delta h|}$ to an overdamped mode with damping rate $\gamma_{\rm OD}\sim|\delta h|/\Gamma^2$; the two scales meet at $|\delta h|\sim\Gamma^4$, i.e. $|\delta h|\sim\Gamma(\Gamma/J)^3$ upon reinstating $J$.

\paragraph*{Quantum-critical Fan and Effective Temperature.---} 
Finite temperature renders $\Pi^K(\omega=0)$ nonzero in Eq.~\eqref{eq:fluct_integral}, changing the logarithmic divergence to the power law $N\langle\delta\phi^2\rangle\sim |T-T_c|^{-1}$. Near the QCP ($T_c\to 0$), the divergence of $N\langle\delta\phi^2\rangle$ organizes the surrounding parameter space into the regions familiar from quantum criticality~\cite{FrerotRoscilde2019}. We focus on the disordered side $\delta h>0$, where the diagram splits into quantum-critical and quantum-disordered regions, sketched in Fig.~\ref{fig:main}(c), which also displays the mirrored structure on the ordered side $\delta h <0$. The supporting numerics are presented in the End Matter.

At low $T$, the dot is in the quantum-disordered regime described by the scaling function $g(\delta h/\Gamma^4)$, including its overdamped--underdamped split. The soft-mode scale interpolates between $\omega_0\sim\sqrt{\delta h}$ (underdamped) and $\gamma_{\rm OD}\sim\delta h/\Gamma^2$ (overdamped), and the two coincide at $\delta h\sim\Gamma^4$. Once $T$ exceeds this scale, the fluctuations become thermally controlled, $N\langle\delta\phi^2\rangle\sim T/a$. The mass $a$ carries a thermal correction $a(T,h)=a_0\delta h+ a_1 T^2$, where $a_1\propto\Gamma$. While $\delta h\gg\Gamma T^2$, one obtains the crossover regime $N\langle\delta\phi^2\rangle\sim T/\delta h$. Once $\Gamma T^2\gg \delta h$, the detuning drops out and the quantum-critical form $N\langle\delta\phi^2\rangle\sim1/(\Gamma T)$ emerges. Because this onset scales as $\sqrt{\delta h/\Gamma}$, stronger dissipation remarkably widens the quantum-critical region rather than washing it out.

Voltage bias produces the same structure in the Gaussian critical sector. Just as with temperature, any finite $V$ gives a finite $\Pi^K(\omega=0)$, while $2-\Pi^R(0)\propto V-V_c$, so the static fluctuations again diverge with the power law $N\langle\delta\phi^2\rangle\sim1/(V-V_c)$~\cite{Mitra_2006, Mitra_2008}. The relevant low-frequency FDR can be parametrized by an effective temperature,
\begin{eqnarray}
T_{\rm eff}(T,V)&=&\frac{1}{2}\left[\lim_{\omega\to0}\frac{\partial}{\partial\omega}
\left(\frac{\Pi^R-\Pi^A}{\Pi^K}\right)\right]^{-1} \label{eq:teff} \\
&=& \;\frac{T}{2}+\frac{V}{4}\coth\left(\!\frac{V}{2T}\right)\left[1+\mathcal{O}\left(\frac{V^2}{h^2}\right)\right],
\label{eq:teff_smallV}
\end{eqnarray}
where the second line is the small-bias expansion, valid for $V\ll h$. In this regime,  Eq.~(\ref{eq:teff_smallV}) interpolates between the limits $T_{\rm eff}\to T$ at $V\ll T$ and $T_{\rm eff}\to V/4$ at $V\gg T$. This is the precise sense in which voltage acts as temperature: $T_{\rm eff}$ fixes the low-frequency quadratic Keldysh action of the soft mode.
Within this domain, replacing $T$ by $T_{\rm eff}$ carries the quantum-critical fan to the driven regime, as confirmed by numerical integration of Eq.~\eqref{eq:fluct_integral} in the End Matter. This effective temperature is measurable by weakly coupling the order parameter to a harmonic probe ~\cite{afonso2026criticalchargecurrentfluctuations}.

\begin{figure}
    \centering
    \includegraphics[width=0.94\linewidth]{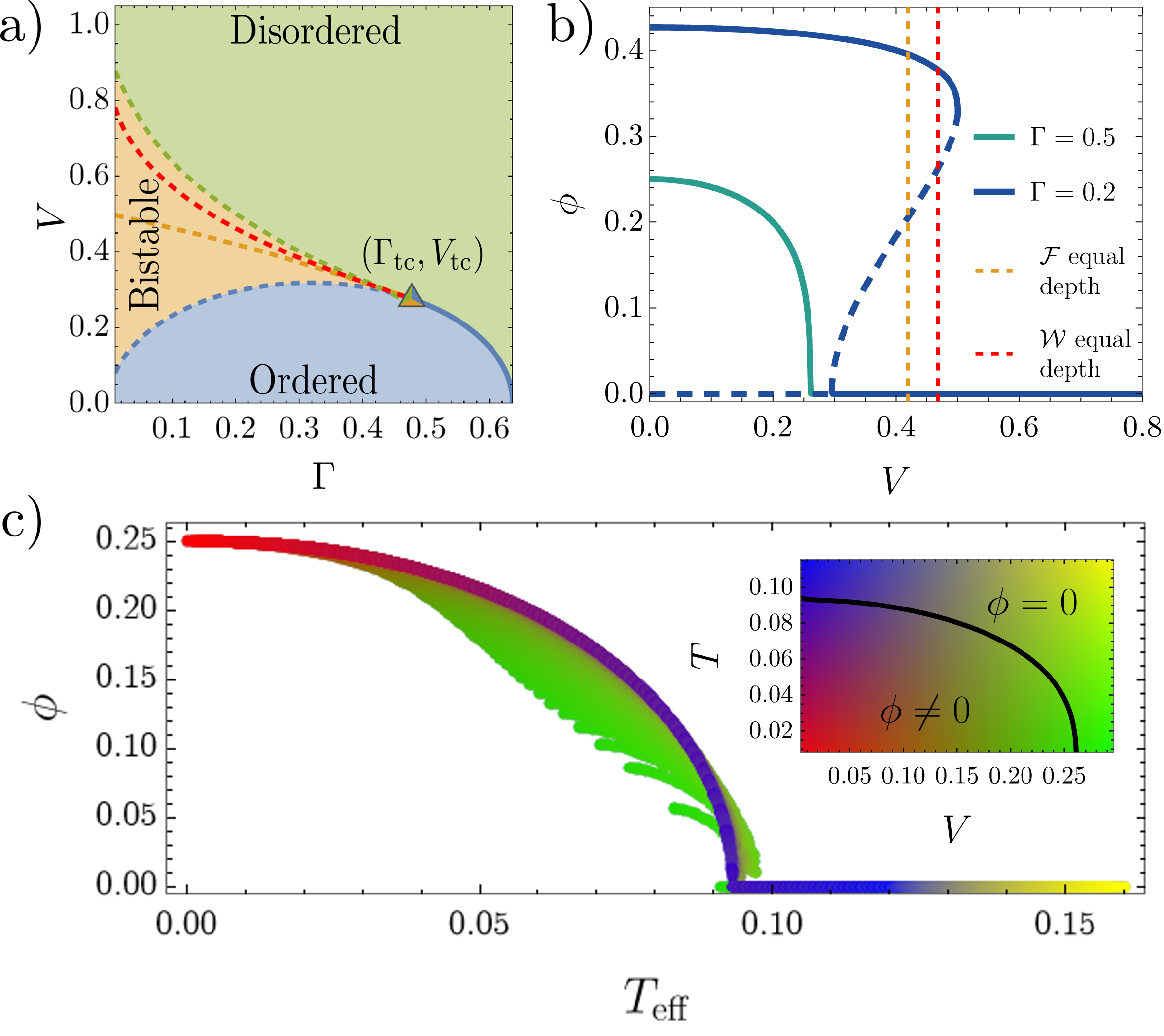}
    \caption{\textbf{Bias-driven transitions} (at $T=h=0$ unless noted otherwise).
(a) Steady-state phase diagram in the $(\Gamma,V)$ plane: an ordered region ($\phi\neq0$ only),
a disordered region ($\phi=0$ only), and a bistable region where $\phi=0$ and $\phi\neq0$ are
simultaneously stable. In the bistable regime, the orange dashed line separates the region where
the global minimum of $\mathcal F$ is disordered (above it) from that where it is ordered (below it). In red, we mark the equal-depth curve of $\mathcal W$ instead. Regions meet at the
tricritical point $(\Gamma_{\rm tc},V_{\rm tc})=(3,\sqrt3)/(2\pi)$.
(b) $\phi(V)$ for $\Gamma=0.5$ (continuous transition) and $\Gamma=0.2$ (first-order transition),
including all solutions of Eq. (\ref{eqn: t0 neq saddle-point}); dashed segments are unstable.
Vertical dashed lines mark the equal-depth points of $\mathcal F$ and $\mathcal W$ at $\Gamma=0.2$.
(c) $\phi$ as a function of $T_{\rm eff}$ (evaluated from the exact FDR definition,
Eq.~\eqref{eq:teff}) for different values of $T$ and $V$ at fixed $\Gamma=0.5$. The inset shows
the color coding in the $(V,T)$ plane.}
    \label{fig:new_phase_1}
    
\end{figure}

\begin{figure*}
    \includegraphics[width=0.99\textwidth]{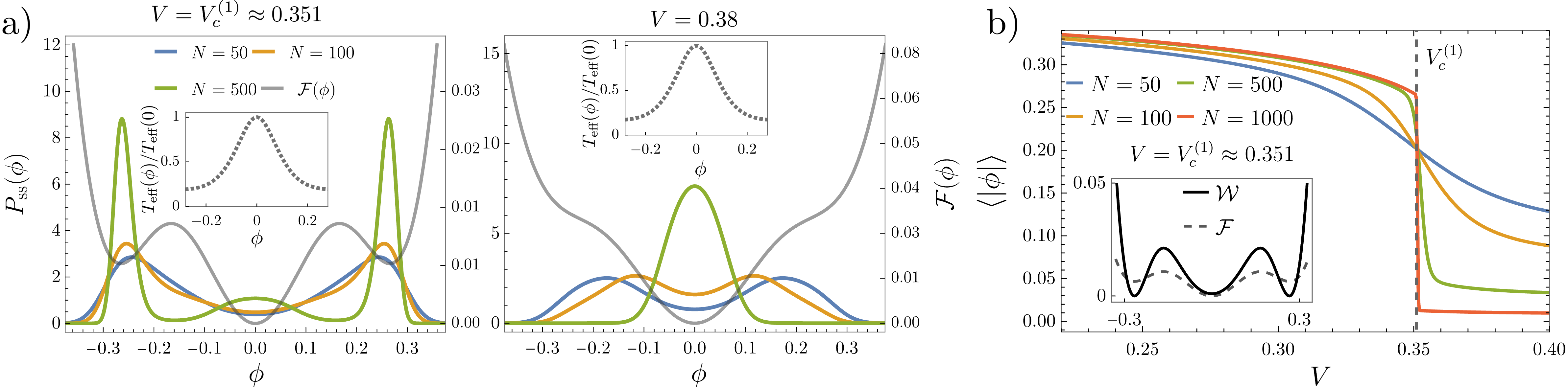}
    \caption[Non-Boltzmann state selection]{\textbf{{Non-Boltzmann state selection}} ($T=h=0$, $\Gamma=0.35$).  \textbf{(a)} Stationary distributions contrasted with $\mathcal{F}[\phi]$ for different $N$ across the bistable region. {$T_{\rm eff}(\phi)/T_\text{eff}(0)$} is plotted on the inset. \textbf{(b)} Expected value $\langle|\phi|\rangle$ for different $N$ across the bistable region. The critical point converges to {$V_c^{(1)}\approx 0.351$}, the equal-depth point of $\mathcal{W}$. In the inset, we plot $\mathcal{F}(\phi)$ and $\mathcal{W}(\phi)$ at $V_c^{(1)}$. At $V=V_c^{(1)}$, $\gamma_\text{OD}(\phi^*)/\Gamma \approx 0.05 $ and $\gamma_\text{OD}(\phi^*)/T_\text{eff}(\phi^*) \approx 0.18$, with $\phi^*$ the mid-point between local minima of $\mathcal{W}$.}
     \label{fig:new_phase_pdf}

\end{figure*}

\paragraph*{{Non-Boltzmann State Selection in the Bistable Regime.}---} 
We now follow the transition driven by the bias itself. At $T=h=0$, the saddle-point equation~\eqref{eq:saddle_general} admits the closed form 
\begin{equation}
\phi=\frac{1}{2\pi}\left[\arctan\left(\frac{V+2\phi}{\Gamma}\right)-\arctan\left(\frac{V-2\phi}{\Gamma}\right)\right],\label{eqn: t0 neq saddle-point}
\end{equation}
whose solution structure changes with $\Gamma$ [Fig.~\ref{fig:new_phase_1}(a,b)]. For large hybridization, the bias-driven transition is continuous, with $V_c=\sqrt{2\Gamma/\pi-\Gamma^2}$; for smaller $\Gamma$ it becomes first order, the two regimes meeting at a tricritical point $(\Gamma_{\rm tc},V_{\rm tc})=(3,\sqrt3)/(2\pi)$. For $\Gamma<\Gamma_\text{tc}$ the self-consistent equation can admit two positive solutions.
{Already at mean-field level the order parameter across the $(T,V)$ plane collapses against $T_{\rm eff}$ near the continuous transition and departs from it once $V$ dominates [Fig.~\ref{fig:new_phase_1}(c)].}

To access fluctuations beyond the Gaussian level, crucial for state selection across the first-order region, we map the Keldysh action onto a stochastic equation for the order parameter. We expand $S[\bm\phi]$ up to quadratic order in $\phi^q$ while keeping all orders of $\phi^{cl}$ intact. The drift term, evaluated for a static field, is exact and integrates to an effective potential $\mathcal F[\phi]$
whose stationary condition $\delta\mathcal F/\delta\phi=0$
reproduces the $N\to\infty$
self-consistent equation~\eqref{eq:saddle_general}. $\mathcal F$ encodes the mean-field phase diagram and classifies the stable and unstable branches shown in Fig.~\ref{fig:new_phase_1}(a,b). Close to the continuous transition, the collective mode is slow, with overdamped relaxation $\gamma_\text{OD}= V_c(V-V_c)/\Gamma+\mathcal{O}(V-V_c)^2$ ($a/\lambda$ of Eq. (\ref{eq:lowfreq_expansion}) at finite $V$ and $\Gamma$). This critical slowing down allows $\phi(t)$ to be treated as quasi-static inside the fermionic trace. A gradient expansion then yields purely relaxational Model-A--type dynamics~\cite{HohenbergHalperin1977},
\begin{equation}
{\lambda[\phi]}\, \dot\phi=-\frac{\delta\mathcal F}{\delta\phi}+\frac1{\sqrt{N}}\,f[\phi(t),t],
\quad
f=\sqrt{-i\Pi^K_0[\phi(t)]/2}\,\xi(t),
\label{eq:langevin_full}
\end{equation}
where $\Pi^K_0[\phi]$ is $\Pi^K(\omega=0)$ evaluated with a quasi-static $\phi(t)$ instead of a static saddle-point $\phi_0$, and $\xi$ is Gaussian noise {with $\langle\xi(t)\xi(t')\rangle=\delta(t-t')$. Likewise, $\lambda[\phi]=-i\left.\partial_\omega\Pi^R[\phi]\right|_{\omega=0}$ in agreement with Eq. (\ref{eq:lowfreq_expansion})}. {The deterministic (noise-independent) potential $\mathcal{F}$ itself has an equilibrium-looking decomposition, derived in the End Matter. It is not a thermodynamic free energy for the driven steady state, but reduces to one in equilibrium.}

A remarkable nonequilibrium signature appears in the stationary distribution. Mapping Eq.~\eqref{eq:langevin_full} to a Fokker--Planck equation, this is readily solved~\cite{risken1996fokker} and yields the following probability distribution:
\begin{equation}
P_{\rm ss}(\phi)\propto\frac{\lambda(\phi)}{T_{\rm eff}(\phi)}\,e^{-N\mathcal W(\phi)},\quad
\mathcal W(\phi)=\int_0^\phi \mathrm{d}\phi'\,
\frac{\mathcal F'(\phi')}{T_{\rm eff}(\phi')},
\label{eq:quasipotential}
\end{equation}
Here $T_{\rm eff}(\phi)=-i\Pi^K_0(\phi)/4\lambda(\phi)$ is the effective temperature of Eq.~\eqref{eq:teff} itself, evaluated in a static background field $\phi$ rather than at the saddle point, and $\mathcal W$ is the associated large-deviation function~\cite{Derrida2007Jul, Touchette2009Jul, Touchette2013Feb, Derrida2025Oct}. The quantity organizing the Gaussian sector therefore also selects the steady state, but sampled along the path connecting the competing states rather than merely at the saddle point, and $\mathcal W$ is a Boltzmann weight of $\mathcal F$ only if $T_{\rm eff}$ is constant along it, which is the case for $V\to0$ where $T_\text{eff}(\phi)\to T$. Out of equilibrium, state selection is then governed by a non-Boltzmann rule that displaces the first-order transition from the equal-depth minima of $\mathcal F$ [Fig.~\ref{fig:new_phase_1}(a), Fig.~\ref{fig:new_phase_pdf}(a,b)], as detailed in the End Matter. Thus the effective temperature complies with the Landauer blowtorch theorem~\cite{Landauer1975, Landauer1988Oct}: although $T_{\rm eff}$ still drives the local fluctuation dynamics, effective thermal behavior is lost because $T_{\rm eff}$ can no longer be considered a global variable of the order parameter.

\paragraph*{Discussion.---}
This work brings within a single controlled framework three fundamental aspects of driven-dissipative quantum criticality: dissipation reshaping the critical point itself~\cite{Yin2014, Vacari_2019_PhysRevA.100.052108, Rossini_2021}, drive acting as an effective temperature~\cite{Mitra_2006, Mitra_2008, Ribeiro2013, Ribeiro2015}, generally dependent on the order parameter~\cite{AronChamon2020, KotliarOOE}, and, at strong drive, genuinely nonequilibrium state selection. At a second-order transition, critical fluctuations probe only a local region of the order-parameter landscape, so a single $T_{\rm eff}$ suffices and the transition retains an effectively thermal character. State selection near a first-order transition is inherently nonlocal in that landscape: it depends on $T_{\rm eff}(\phi)$ along the entire path connecting competing states. In equilibrium, $T_{\rm eff}(\phi)=T$ is constant, and the relative depths of the minima of $\mathcal F$ determine their stationary weights. Driving removes this constraint, so variations of $T_{\rm eff}$ across the landscape generically decouple stationary-state selection from the deterministic potential. Since drive-induced multistability and first-order transitions~\cite{KotliarOOE, DallaTorre2010, Fink_2017, Minganti2018_SpectralLiouvillian, Walldorf2019Sep, oliveira2024r, Debecker2025} are ubiquitous in nonequilibrium quantum matter, the order-parameter dependence of the FDR provides a simple criterion for distinguishing emergent thermal behavior from genuinely nonequilibrium state selection.

\textbf{Acknowledgements.}
We thank Hans Christiansen for useful discussions. T.J. and P.R. acknowledge support by FCT through Grant No.~UID/04540/2025 (DOI: 10.54499/UID/04540/2025) to the I\&D unit Centro de F\'isica e Engenharia de Materiais Avan\c{c}ados (CeFEMA) and through project SCALE-QLT (DOI: 10.54499/2024.16192.PEX). T.J. acknowledges support from FCT through Grant No. {2025.03249.BD} and under the Project 2026.00107.CPCA.A1 at the Deucalion supercomputer. This work has benefited from discussions and networking activities within COST Action CA24109, \emph{Many-body Open Quantum Systems (QOpen)}, supported by COST (European Cooperation in Science and Technology).

\bibliographystyle{apsrev4-2}
\bibliography{main}\clearpage

@article{Landauer1988Oct,
	author = {Landauer, Rolf},
	title = {{Motion out of noisy states}},
	journal = {J. Stat. Phys.},
	volume = {53},
	number = {1},
	pages = {233--248},
	year = {1988},
	month = oct,
	issn = {1572-9613},
	publisher = {Kluwer Academic Publishers-Plenum Publishers},
	doi = {10.1007/BF01011555}
}

@PREAMBLE{
 "\providecommand{\noopsort}[1]{}" 
 # "\providecommand{\singleletter}[1]{#1}%" 
}

@article{Green2005Dec,
	author = {Green, A. G. and Sondhi, S. L.},
	title = {{Nonlinear Quantum Critical Transport and the Schwinger Mechanism for a Superfluid-Mott-Insulator Transition of Bosons}},
	journal = {Phys. Rev. Lett.},
	volume = {95},
	number = {26},
	pages = {267001},
	year = {2005},
	month = dec,
	publisher = {American Physical Society},
	doi = {10.1103/PhysRevLett.95.267001}
}

@article{Feldman2005Oct,
	author = {Feldman, D. E.},
	title = {{Nonequilibrium Quantum Phase Transition in Itinerant Electron Systems}},
	journal = {Phys. Rev. Lett.},
	volume = {95},
	number = {17},
	pages = {177201},
	year = {2005},
	month = oct,
	publisher = {American Physical Society},
	doi = {10.1103/PhysRevLett.95.177201}
}

@article{Dalidovich2004Jul,
	author = {Dalidovich, Denis and Phillips, Philip},
	title = {{Nonlinear Transport near a Quantum Phase Transition in Two Dimensions}},
	journal = {Phys. Rev. Lett.},
	volume = {93},
	number = {2},
	pages = {027004},
	year = {2004},
	month = jul,
	publisher = {American Physical Society},
	doi = {10.1103/PhysRevLett.93.027004}
}

@article{Kawamura2025Dec,
	author = {Kawamura, Taira and Ohashi, Yoji},
	title = {{Engineering Nonequilibrium Superconducting Phases in a Voltage-Driven Superconductor Under an External Magnetic Field}},
	journal = {Ann. Phys.},
	volume = {537},
	number = {12},
	pages = {e2500102},
	year = {2025},
	month = dec,
	issn = {0003-3804},
	publisher = {John Wiley {\&} Sons, Ltd},
	doi = {10.1002/andp.202500102}
}

@article{Rota2019Mar,
	author = {Rota, Riccardo and Minganti, Fabrizio and Ciuti, Cristiano and Savona, Vincenzo},
	title = {{Quantum Critical Regime in a Quadratically Driven Nonlinear Photonic Lattice}},
	journal = {Phys. Rev. Lett.},
	volume = {122},
	number = {11},
	pages = {110405},
	year = {2019},
	month = mar,
	publisher = {American Physical Society},
	doi = {10.1103/PhysRevLett.122.110405}
}

@article{Derrida2025Oct,
	author = {Derrida, Bernard},
	title = {{Lecture notes on large deviations in non-equilibrium diffusive systems}},
	journal = {SciPost Phys. Lect. Notes},
	pages = {106},
	year = {2025},
	month = oct,
	issn = {2590-1990},
	doi = {10.21468/SciPostPhysLectNotes.106}
}

@article{Derrida2007Jul,
	author = {Derrida, Bernard},
	title = {{Non-equilibrium steady states: fluctuations and large deviations of the density and of the current}},
	journal = {J. Stat. Mech.: Theory Exp.},
	volume = {2007},
	number = {07},
	pages = {P07023},
	year = {2007},
	month = jul,
	issn = {1742-5468},
	publisher = {IOP Publishing},
	doi = {10.1088/1742-5468/2007/07/P07023}
}

@incollection{Touchette2013Feb,
	author = {Touchette, Hugo and Harris, Rosemary J.},
	title = {{Large Deviation Approach to Nonequilibrium Systems}},
	booktitle = {{Nonequilibrium Statistical Physics of Small Systems}},
	journal = {Wiley Online Library},
	pages = {335--360},
	year = {2013},
	month = feb,
	isbn = {978-3-52765870-1},
	publisher = {Wiley-VCH Verlag GmbH {\&} Co. KGaA},
	address = {Weinheim, Germany},
	doi = {10.1002/9783527658701.ch11}
}

@article{Touchette2009Jul,
	author = {Touchette, Hugo},
	title = {{The large deviation approach to statistical mechanics}},
	journal = {Phys. Rep.},
	volume = {478},
	number = {1},
	pages = {1--69},
	year = {2009},
	month = jul,
	issn = {0370-1573},
	publisher = {North-Holland},
	doi = {10.1016/j.physrep.2009.05.002}
}

@article{Walldorf2019Sep,
	author = {Walldorf, Nicklas and Kennes, Dante M. and Paaske, Jens and Millis, Andrew J.},
	title = {{The antiferromagnetic phase of the Floquet-driven Hubbard model}},
	journal = {Phys. Rev. B},
	volume = {100},
	number = {12},
	pages = {121110},
	year = {2019},
	month = sep,
	publisher = {American Physical Society},
	doi = {10.1103/PhysRevB.100.121110}
}

@article{Chung2009May,
	author = {Chung, Chung-Hou and Le Hur, Karyn and Vojta, Matthias and W{\ifmmode\ddot{o}\else\"{o}\fi}lfle, Peter},
	title = {{Nonequilibrium Transport at a Dissipative Quantum Phase Transition}},
	journal = {Phys. Rev. Lett.},
	volume = {102},
	number = {21},
	pages = {216803},
	year = {2009},
	month = may,
	publisher = {American Physical Society},
	doi = {10.1103/PhysRevLett.102.216803}
}

@article{Sela2009Jan,
	author = {Sela, Eran and Affleck, Ian},
	title = {{Nonequilibrium Transport through Double Quantum Dots: Exact Results near a Quantum Critical Point}},
	journal = {Phys. Rev. Lett.},
	volume = {102},
	number = {4},
	pages = {047201},
	year = {2009},
	month = jan,
	publisher = {American Physical Society},
	doi = {10.1103/PhysRevLett.102.047201}
}

@article{Kirchner2009Nov,
	author = {Kirchner, Stefan and Si, Qimiao},
	title = {{Quantum Criticality Out of Equilibrium: Steady State in a Magnetic Single-Electron Transistor}},
	journal = {Phys. Rev. Lett.},
	volume = {103},
	number = {20},
	pages = {206401},
	year = {2009},
	month = nov,
	publisher = {American Physical Society},
	doi = {10.1103/PhysRevLett.103.206401}
}

@article{Zhang2021Feb,
	author = {Zhang, Gu and Chung, Chung-Hou and Ke, Chung-Ting and Lin, Chao-Yun and Mebrahtu, Henok and Smirnov, Alex I. and Finkelstein, Gleb and Baranger, Harold U.},
	title = {{Nonequilibrium quantum critical steady state: Transport through a dissipative resonant level}},
	journal = {Phys. Rev. Res.},
	volume = {3},
	number = {1},
	pages = {013136},
	year = {2021},
	month = feb,
	publisher = {American Physical Society},
	doi = {10.1103/PhysRevResearch.3.013136}
}

@article{Maghrebi2016Jan,
	author = {Maghrebi, Mohammad F. and Gorshkov, Alexey V.},
	title = {{Nonequilibrium many-body steady states via Keldysh formalism}},
	journal = {Phys. Rev. B},
	volume = {93},
	number = {1},
	pages = {014307},
	year = {2016},
	month = jan,
	publisher = {American Physical Society},
	doi = {10.1103/PhysRevB.93.014307}
}

@article{Lundgren2020Sep,
	author = {Lundgren, Rex and Gorshkov, Alexey V. and Maghrebi, Mohammad F.},
	title = {{Nature of the nonequilibrium phase transition in the non-Markovian driven Dicke model}},
	journal = {Phys. Rev. A},
	volume = {102},
	number = {3},
	pages = {032218},
	year = {2020},
	month = sep,
	publisher = {American Physical Society},
	doi = {10.1103/PhysRevA.102.032218}
}

@article{Ribeiro2013,
  title = {Local quantum criticality out of equilibrium: effective temperatures and scaling in the steady state regime},
  author = {Ribeiro, P. and Si, Q. and Kirchner, S.},
  journal = {EPL},
  volume = {102},
  pages = {50001},
  year = {2013}
}

@article{Ribeiro2015,
  title = {Steady-state dynamics and effective temperatures of quantum criticality in an open system},
  author = {Ribeiro, P. and Zamani, F. and Kirchner, S.},
  journal = {Phys. Rev. Lett.},
  volume = {115},
  pages = {220602},
  year = {2015}
}

@article{Dusuel2004,
  author  = {Dusuel, S. and Vidal, J.},
  title   = {Finite-size scaling exponents of the Lipkin-Meshkov-Glick model},
  journal = {Physical Review Letters},
  volume  = {93},
  pages   = {237204},
  year    = {2004},
  doi     = {10.1103/PhysRevLett.93.237204}
}

@article{Yin2014,
  author  = {Yin, S. and Mai, P. and Zhong, F.},
  title   = {Nonequilibrium quantum criticality in open systems},
  journal = {Physical Review B},
  volume  = {89},
  pages   = {094108},
  year    = {2014},
  doi     = {10.1103/PhysRevB.89.094108}
}

@article{Sieberer2013,
  author  = {Sieberer, L. M. and Huber, S. D. and Altman, E. and Diehl, S.},
  title   = {Dynamical critical phenomena in driven-dissipative systems},
  journal = {Physical Review Letters},
  volume  = {110},
  pages   = {195301},
  year    = {2013},
  doi     = {10.1103/PhysRevLett.110.195301}
}

@article{DallaTorre2013,
  author  = {Dalla Torre, E. G. and Diehl, S. and Lukin, M. D. and Sachdev, S. and Strack, P.},
  title   = {Keldysh approach for nonequilibrium phase transitions in quantum optics: Beyond the Dicke model in optical cavities},
  journal = {Physical Review A},
  volume  = {87},
  pages   = {023831},
  year    = {2013},
  doi     = {10.1103/PhysRevA.87.023831}
}

@article{HohenbergHalperin1977,
  author  = {Hohenberg, P. C. and Halperin, B. I.},
  title   = {Theory of dynamic critical phenomena},
  journal = {Reviews of Modern Physics},
  volume  = {49},
  pages   = {435--479},
  year    = {1977},
  doi     = {10.1103/RevModPhys.49.435}
}

@article{FrerotRoscilde2019,
  title   = {Reconstructing the quantum critical fan of strongly correlated
             systems using quantum correlations},
  author  = {Fr{\'e}rot, Ir{\'e}n{\'e}e and Roscilde, Tommaso},
  journal = {Nat. Commun.},
  volume  = {10},
  pages   = {577},
  year    = {2019},
  doi     = {10.1038/s41467-019-08324-9},
  eprint  = {1805.03140},
  archivePrefix = {arXiv},
  primaryClass  = {cond-mat.str-el},
}

@article{KopylovSchaller2019,
  title   = {Polaron-transformed dissipative Lipkin-Meshkov-Glick model},
  author  = {Kopylov, Wassilij and Schaller, Gernot},
  journal = {Phys. Rev. A},
  volume  = {100},
  issue   = {6},
  pages   = {063815},
  year    = {2019},
  doi     = {10.1103/PhysRevA.100.063815},
  eprint  = {1906.04260},
  archivePrefix = {arXiv},
  primaryClass  = {quant-ph},
}

@misc{oliveira2024r,
      title={Voltage-Driven Breakdown of Electronic Order}, 
      author={Miguel M. Oliveira and Pedro Ribeiro and Stefan Kirchner},
      year={2024},
      eprint={2405.09512},
      archivePrefix={arXiv},
      primaryClass={cond-mat.str-el},
      url={https://arxiv.org/abs/2405.09512}, 
}

@book{SachdevQPT,
   title =     {Quantum Phase Transitions},
   author =    {Subir Sachdev},
   publisher = {Cambridge University Press},
   isbn =      {0521514681; 9780521514682},
   year =      {2011},
   edition =   {2},
}

@article{Bode2012Mar,
	author = {Bode, Niels and Arrachea, Liliana and Lozano, Gustavo S. and Nunner, Tamara S. and von Oppen, Felix},
	title = {{Current-induced switching in transport through anisotropic magnetic molecules}},
	journal = {Phys. Rev. B},
	volume = {85},
	number = {11},
	pages = {115440},
	year = {2012},
	month = mar,
	publisher = {American Physical Society},
	doi = {10.1103/PhysRevB.85.115440}
}

@article{Gorokhov2004Apr,
	author = {Gorokhov, Denis A. and Brouwer, Piet W.},
	title = {{Combined effect of electron-electron interactions and spin-orbit scattering in metal nanoparticles}},
	journal = {Phys. Rev. B},
	volume = {69},
	number = {15},
	pages = {155417},
	year = {2004},
	month = apr,
	publisher = {American Physical Society},
	doi = {10.1103/PhysRevB.69.155417}
}

@article{Basko2009Feb,
	author = {Basko, Denis M. and Vavilov, Maxim G.},
	title = {{Stochastic dynamics of magnetization in a ferromagnetic nanoparticle out of equilibrium}},
	journal = {Phys. Rev. B},
	volume = {79},
	number = {6},
	pages = {064418},
	year = {2009},
	month = feb,
	publisher = {American Physical Society},
	doi = {10.1103/PhysRevB.79.064418}
}

@article{Waintal2003Dec,
	author = {Waintal, Xavier and Brouwer, Piet W.},
	title = {{Tunable Magnetic Relaxation Mechanism in Magnetic Nanoparticles}},
	journal = {Phys. Rev. Lett.},
	volume = {91},
	number = {24},
	pages = {247201},
	year = {2003},
	month = dec,
	publisher = {American Physical Society},
	doi = {10.1103/PhysRevLett.91.247201}
}

@article{Brouwer1999Nov,
	author = {Brouwer, P. W. and Oreg, Yuval and Halperin, B. I.},
	title = {{Mesoscopic fluctuations of the ground-state spin of a small metal particle}},
	journal = {Phys. Rev. B},
	volume = {60},
	number = {20},
	pages = {R13977--R13980},
	year = {1999},
	month = nov,
	publisher = {American Physical Society},
	doi = {10.1103/PhysRevB.60.R13977}
}

@article{Andreev1998Oct,
	author = {Andreev, A. V. and Kamenev, A.},
	title = {{Itinerant Ferromagnetism in Disordered Metals: A Mean-Field Theory}},
	journal = {Phys. Rev. Lett.},
	volume = {81},
	number = {15},
	pages = {3199--3202},
	year = {1998},
	month = oct,
	publisher = {American Physical Society},
	doi = {10.1103/PhysRevLett.81.3199}
}

@article{Burmistrov2020May,
	author = {Burmistrov, I. S. and Gefen, Y. and Shapiro, D. S. and Shnirman, A.},
	title = {{Mesoscopic Stoner Instability in Open Quantum Dots: Suppression of Coleman-Weinberg Mechanism by Electron Tunneling}},
	journal = {Phys. Rev. Lett.},
	volume = {124},
	number = {19},
	pages = {196801},
	year = {2020},
	month = may,
	publisher = {American Physical Society},
	doi = {10.1103/PhysRevLett.124.196801}
}

@article{Sharafutdinov2014Nov,
	author = {Sharafutdinov, A. U. and Lyubshin, D. S. and Burmistrov, I. S.},
	title = {{Spin fluctuations in quantum dots}},
	journal = {Phys. Rev. B},
	volume = {90},
	number = {19},
	pages = {195308},
	year = {2014},
	month = nov,
	publisher = {American Physical Society},
	doi = {10.1103/PhysRevB.90.195308}
}

@article{Sieberer_2016,
   title={Keldysh field theory for driven open quantum systems},
   volume={79},
   ISSN={1361-6633},
   url={http://dx.doi.org/10.1088/0034-4885/79/9/096001},
   number={9},
   journal={Reports on Progress in Physics},
   publisher={IOP Publishing},
   author={Sieberer, L M and Buchhold, M and Diehl, S},
   year={2016},
   month=aug, pages={096001} }

@book{Kamenev2023Jan,
	author = {Kamenev, Alex},
	title = {{Field Theory of Non-Equilibrium Systems}},
    edition   = {2},
	journal = {Cambridge Core},
	year = {2023},
	month = jan,
	isbn = {978-1-10876926-6},
	publisher = {Cambridge University Press},
	address = {Cambridge, England, UK},
	doi = {10.1017/9781108769266}
}

@article{Coleman_2005_Criticality,
   title={Quantum criticality},
   volume={433},
   ISSN={1476-4687},
   url={http://dx.doi.org/10.1038/nature03279},
   DOI={10.1038/nature03279},
   number={7023},
   journal={Nature},
   publisher={Springer Science and Business Media LLC},
   author={Coleman, Piers and Schofield, Andrew J.},
   year={2005},
   month=jan, pages={226–229} }

@article{Alhassid_2000_QD_StatisticalT,
   title={The statistical theory of quantum dots},
   volume={72},
   ISSN={1539-0756},
   url={http://dx.doi.org/10.1103/RevModPhys.72.895},
   DOI={10.1103/revmodphys.72.895},
   number={4},
   journal={Reviews of Modern Physics},
   publisher={American Physical Society (APS)},
   author={Alhassid, Y.},
   year={2000},
   month=oct, pages={895–968} }

@article{UniversalQDHamiltonian2,
  title = {Mesoscopic magnetization fluctuations for metallic grains close to the Stoner instability},
  author = {Kurland, I. L. and Aleiner, I. L. and Altshuler, B. L.},
  journal = {Phys. Rev. B},
  volume = {62},
  issue = {22},
  pages = {14886--14897},
  numpages = {0},
  year = {2000},
  month = {Dec},
  publisher = {American Physical Society},
  doi = {10.1103/PhysRevB.62.14886},
  url = {https://link.aps.org/doi/10.1103/PhysRevB.62.14886}
}

@article{LIPKIN1965,
  author  = {Lipkin, H. J. and Meshkov, N. and Glick, A. J.},
  title   = {Validity of many-body approximation methods for a solvable model: (I). Exact solutions and perturbation theory},
  journal = {Nuclear Physics},
  volume  = {62},
  number  = {2},
  pages   = {188--198},
  year    = {1965},
  doi     = {10.1016/0029-5582(65)90862-X}
}

@article{Garanin_1998_SingleMoleculeMagnets,
   title={Quantum-classical transition of the escape rate of a uniaxial spin system in an arbitrarily directed field},
   volume={57},
   ISSN={1095-3795},
   url={http://dx.doi.org/10.1103/PhysRevB.57.13639},
   DOI={10.1103/physrevb.57.13639},
   number={21},
   journal={Physical Review B},
   publisher={American Physical Society (APS)},
   author={Garanin, D. A. and Martínez Hidalgo, X. and Chudnovsky, E. M.},
   year={1998},
   month=jun, pages={13639–13654} }

@article{LMG_BEC,
  title = {Quantum superposition states of Bose-Einstein condensates},
  author = {Cirac, J. I. and Lewenstein, M. and M\o{}lmer, K. and Zoller, P.},
  journal = {Phys. Rev. A},
  volume = {57},
  issue = {2},
  pages = {1208--1218},
  numpages = {0},
  year = {1998},
  month = {Feb},
  publisher = {American Physical Society},
  doi = {10.1103/PhysRevA.57.1208},
  url = {https://link.aps.org/doi/10.1103/PhysRevA.57.1208}
}

@article{CavityQED_LMG,
  title = {Dynamical Quantum Phase Transitions in the Dissipative Lipkin-Meshkov-Glick Model with Proposed Realization in Optical Cavity QED},
  author = {Morrison, S. and Parkins, A. S.},
  journal = {Phys. Rev. Lett.},
  volume = {100},
  issue = {4},
  pages = {040403},
  numpages = {4},
  year = {2008},
  month = {Jan},
  publisher = {American Physical Society},
  doi = {10.1103/PhysRevLett.100.040403},
  url = {https://link.aps.org/doi/10.1103/PhysRevLett.100.040403}
}

@article{FerreiraRibeiro2019_LMG_Dissipation,
  title        = {Lipkin--Meshkov--Glick model with Markovian dissipation: Classical bifurcations and quantum fluctuations},
  author       = {Ferreira, João S. and Ribeiro, Pedro},
  journal      = {Physical Review B},
  volume       = {100},
  number       = {18},
  pages        = {184422},
  year         = {2019},
  doi          = {10.1103/PhysRevB.100.184422},
  url          = {https://journals.aps.org/prb/abstract/10.1103/PhysRevB.100.184422}
}

@article{Vojta_2003,
   title={Quantum phase transitions},
   volume={66},
   ISSN={1361-6633},
   url={http://dx.doi.org/10.1088/0034-4885/66/12/R01},
   DOI={10.1088/0034-4885/66/12/r01},
   number={12},
   journal={Reports on Progress in Physics},
   publisher={IOP Publishing},
   author={Vojta, Matthias},
   year={2003},
   month=nov, pages={2069–2110} }

@article{Gegenwart2008_HeavyFermionQCP,
  title        = {Quantum criticality in heavy-fermion metals},
  author       = {Gegenwart, Philipp and Si, Qimiao and Steglich, Frank},
  journal      = {Nature Physics},
  volume       = {4},
  pages        = {186--197},
  year         = {2008},
  doi          = {10.1038/nphys892}
}

@article{Keimer2015_HighTcQPT,
  title        = {From quantum matter to high-temperature superconductivity in copper oxides},
  author       = {Keimer, Bernhard and Kivelson, Steven A. and Norman, Michael R. and Uchida, Shinichi and Zaanen, Jan},
  journal      = {Nature},
  volume       = {518},
  pages        = {179--186},
  year         = {2015},
  doi          = {10.1038/nature14165}
}

@book{Dutta2015_QPT_SpinModels,
  title     = {{Quantum Phase Transitions in Transverse Field Spin Models:
                From Statistical Physics to Quantum Information}},
  author    = {Dutta, Amit and Aeppli, Gabriel and Chakrabarti, Bikas K. and
               Divakaran, Uma and Rosenbaum, Thomas F. and Sen, Diptiman},
  publisher = {Cambridge University Press},
  address   = {Cambridge, England},
  year      = {2015},
  isbn      = {9781107068797},
  doi       = {10.1017/CBO9781107706057}
}

@misc{Okugawa2026Jan,
  title         = {{Dynamical Instability in a Floquet-Driven Dissipative System}},
  author        = {Okugawa, Takuya and Paaske, Jens and Eckstein, Martin and
                   Sentef, Michael A. and Rubio, Angel and Millis, Andrew J.},
  year          = {2026},
  eprint        = {2601.04451},
  archivePrefix = {arXiv},
  primaryClass  = {cond-mat.str-el}
}

@article{Fink_2017,
   title={Signatures of a dissipative phase transition in photon correlation measurements},
   volume={14},
   ISSN={1745-2481},
   url={http://dx.doi.org/10.1038/s41567-017-0020-9},
   DOI={10.1038/s41567-017-0020-9},
   number={4},
   journal={Nature Physics},
   publisher={Springer Science and Business Media LLC},
   author={Fink, Thomas and Schade, Anne and Höfling, Sven and Schneider, Christian and Imamoglu, Ataç},
   year={2018},
   month=dec, pages={365–369} }

@article{Kessler2012_SpinDissipation,
  title        = {Dissipative phase transition in a central spin system},
  author       = {Kessler, E. M. and Giedke, G. and Imamoglu, A. and Yelin, S. F. and Lukin, M. D. and Cirac, J. I.},
  journal      = {Physical Review A},
  volume       = {86},
  number       = {1},
  pages        = {012116},
  year         = {2012},
  doi          = {10.1103/PhysRevA.86.012116}
}

@article{Marcuzzi2014_RydbergNoneq,
  title        = {Universal non-equilibrium properties of dissipative Rydberg gases},
  author       = {Marcuzzi, Matteo and Levi, Emanuele and Diehl, Sebastian and Garrahan, Juan P. and Lesanovsky, Igor},
  journal      = {Physical Review Letters},
  volume       = {113},
  number       = {21},
  pages        = {210401},
  year         = {2014},
  doi          = {10.1103/PhysRevLett.113.210401}
}

@book{Schaller2014_OpenQS,
  title        = {Open Quantum Systems Far from Equilibrium},
  author       = {Schaller, Gernot},
  publisher    = {Springer},
  series       = {Lecture Notes in Physics},
  volume       = {881},
  year         = {2014},
  doi          = {10.1007/978-3-319-03877-3}
}

@article{DeVega2017_NonMarkovReview,
  title        = {Dynamics of non-Markovian open quantum systems},
  author       = {de Vega, In\'es and Alonso, Daniel},
  journal      = {Reviews of Modern Physics},
  volume       = {89},
  number       = {1},
  pages        = {015001},
  year         = {2017},
  doi          = {10.1103/RevModPhys.89.015001}
}

@article{Minganti2018_SpectralLiouvillian,
  title        = {Spectral theory of Liouvillians for dissipative phase transitions},
  author       = {Minganti, Fabrizio and Biella, Alessio and Bartolo, Nicola and Ciuti, Cristiano},
  journal      = {Physical Review A},
  volume       = {98},
  number       = {4},
  pages        = {042118},
  year         = {2018},
  doi          = {10.1103/PhysRevA.98.042118}
}

@incollection{GanzhornWernsdorfer2021,
  author    = {Ganzhorn, Marc and Wernsdorfer, Wolfgang},
  title     = {Molecular Quantum Spintronics Using Single-Molecule Magnets},
  booktitle = {Molecular Magnets: Physics and Applications},
  editor    = {Bartolom\'{e}, Juan and Luis, Fernando and Fern\'{a}ndez, Julio F.},
  series    = {NanoScience and Technology},
  pages     = {319--364},
  year      = {2014},
  publisher = {Springer},
  doi       = {10.1007/978-3-642-40609-6_13}
}

@article{BotetJullien1983,
  author    = {R. Botet and R. Jullien},
  title     = {Large-size critical behaviour of infinitely coordinated systems},
  journal   = {Physical Review B},
  volume    = {28},
  number    = {7},
  pages     = {3955--3965},
  year      = {1983},
  doi       = {10.1103/PhysRevB.28.3955}
}

@article{Debecker2025,
  title = {Role of non-Markovian dissipation in quantum phase transitions: Tricriticality, spin squeezing, and directional symmetry breaking},
  author = {Debecker, Baptiste and Pausch, Lukas and Louvet, Jonathan and Bastin, Thierry and Martin, John and Damanet, François},
  journal = {Phys. Rev. A},
  volume = {112},
  issue = {1},
  pages = {012210},
  numpages = {15},
  year = {2025},
  month = {Jul},
  publisher = {American Physical Society},
  doi = {10.1103/4wq2-wyh4},
  url = {https://link.aps.org/doi/10.1103/4wq2-wyh4}
}

@article{chakraborty_sensarma_2018,
  title     = {Power-law tails and non-Markovian dynamics in open quantum systems: An exact solution from Keldysh field theory},
  author    = {Chakraborty, Anirban and Sensarma, Rajdeep},
  journal   = {Physical Review B},
  volume    = {97},
  number    = {10},
  pages     = {104306},
  year      = {2018},
  publisher = {American Physical Society},
  doi       = {10.1103/PhysRevB.97.104306}
}

@article{KotliarOOE, title = {Nonequilibrium mean-field theory of resistive phase transitions}, author = {Han, Jong E. and Li, Jiajun and Aron, Camille and Kotliar, Gabriel}, journal = {Phys. Rev. B}, volume = {98}, issue = {3}, pages = {035145}, numpages = {16}, year = {2018}, month = {Jul}, publisher = {American Physical Society}, doi = {10.1103/PhysRevB.98.035145}, url = {https://link.aps.org/doi/10.1103/PhysRevB.98.035145} }

@article{Mitra_2008,
   title={Current-driven quantum criticality in itinerant electron ferromagnets},
   volume={77},
   ISSN={1550-235X},
   url={http://dx.doi.org/10.1103/PhysRevB.77.220404},
   DOI={10.1103/physrevb.77.220404},
   pages={220404},
   number={22},
   journal={Physical Review B},
   publisher={American Physical Society (APS)},
   author={Mitra, Aditi and Millis, Andrew J.},
   year={2008},
   month=jun }

@article{Mitra_2006,
   title={Nonequilibrium Quantum Criticality in Open Electronic Systems},
   volume={97},
   ISSN={1079-7114},
   url={http://dx.doi.org/10.1103/PhysRevLett.97.236808},
   DOI={10.1103/physrevlett.97.236808},
   pages={236808},
   number={23},
   journal={Physical Review Letters},
   publisher={American Physical Society (APS)},
   author={Mitra, Aditi and Takei, So and Kim, Yong Baek and Millis, A. J.},
   year={2006},
   month=dec }

@misc{afonso2026criticalchargecurrentfluctuations,
      title={Critical Charge and Current Fluctuations across a Voltage-Driven Phase Transition}, 
      author={José F. B. Afonso and Stefan Kirchner and Pedro Ribeiro},
      year={2026},
      eprint={2601.20474},
      archivePrefix={arXiv},
      primaryClass={cond-mat.str-el},
      url={https://arxiv.org/abs/2601.20474}, 
}

@article{Wilms_2012,
   title={Finite-temperature mutual information in a simple phase transition},
   volume={2012},
   ISSN={1742-5468},
   url={http://dx.doi.org/10.1088/1742-5468/2012/01/P01023},
   DOI={10.1088/1742-5468/2012/01/p01023},
   number={01},
   journal={Journal of Statistical Mechanics: Theory and Experiment},
   publisher={IOP Publishing},
   author={Wilms, Johannes and Vidal, Julien and Verstraete, Frank and Dusuel, Sébastien},
   year={2012},
   month=jan, pages={P01023} }

@article{Dusuel_2005,
   title={Continuous unitary transformations and finite-size scaling exponents in the Lipkin-Meshkov-Glick model},
   volume={71},
   ISSN={1550-235X},
   url={http://dx.doi.org/10.1103/PhysRevB.71.224420},
   DOI={10.1103/physrevb.71.224420},
   pages={224420},
   number={22},
   journal={Physical Review B},
   publisher={American Physical Society (APS)},
   author={Dusuel, Sébastien and Vidal, Julien},
   year={2005},
   month=jun }

@book{Bartolome2014Molecular,
  title     = {Molecular Magnets: Physics and Applications},
  editor    = {Bartolomé, Juan and Luis, Fernando and Fernández, Julio F.},
  series    = {NanoScience and Technology},
  publisher = {Springer},
  year      = {2014},
  isbn      = {9783642406089}
}

@article{Vacari_2019_PhysRevA.100.052108,
  title = {Competing coherent and dissipative dynamics close to quantum criticality},
  author = {Nigro, Davide and Rossini, Davide and Vicari, Ettore},
  journal = {Phys. Rev. A},
  volume = {100},
  issue = {5},
  pages = {052108},
  numpages = {9},
  year = {2019},
  month = {Nov},
  publisher = {American Physical Society},
  doi = {10.1103/PhysRevA.100.052108},
  url = {https://link.aps.org/doi/10.1103/PhysRevA.100.052108}
}

@article{Rossini_2021,
   title={Coherent and dissipative dynamics at quantum phase transitions},
   volume={936},
   ISSN={0370-1573},
   url={http://dx.doi.org/10.1016/j.physrep.2021.08.003},
   DOI={10.1016/j.physrep.2021.08.003},
   journal={Physics Reports},
   publisher={Elsevier BV},
   author={Rossini, Davide and Vicari, Ettore},
   year={2021},
   month=Nov, pages={1–110} }

@misc{hooley2013LMGquasilocalquantum,
      title={The {L}ipkin-{M}eshkov-{G}lick model: 'quasi-local' quantum criticality in nuclear physics}, 
      author={C. A. Hooley and P. D. Stevenson},
      year={2011},
      eprint={1102.1583},
      archivePrefix={arXiv},
      primaryClass={nucl-th},
      url={https://arxiv.org/abs/1102.1583}, 
}

@article{Cugliandolo1997_EffectiveTemperatures,
  author  = {Cugliandolo, L. F. and Kurchan, J. and Peliti, L.},
  title   = {Energy flow, partial equilibration, and effective temperatures in systems with slow dynamics},
  journal = {Phys. Rev. E},
  volume  = {55},
  issue   = {4},
  pages   = {3898--3914},
  year    = {1997},
  doi     = {10.1103/PhysRevE.55.3898}
}

@article{SiebererDiehl2025,
  author  = {Sieberer, L. M. and Buchhold, M. and Marino, J. and Diehl, S.},
  title   = {Universality in driven open quantum matter},
  journal = {Rev. Mod. Phys.},
  volume  = {97},
  issue   = {2},
  pages   = {025004},
  year    = {2025},
  doi     = {10.1103/RevModPhys.97.025004}
}

@article{DallaTorre2010,
  author  = {Dalla Torre, Emanuele G. and Demler, Eugene and Giamarchi, Thierry and Altman, Ehud},
  title   = {Quantum critical states and phase transitions in the presence of non-equilibrium noise},
  journal = {Nature Physics},
  volume  = {6},
  pages   = {806--810},
  year    = {2010},
  doi     = {10.1038/nphys1754}
}

@book{tauber2014critical,
  author    = {T{\"a}uber, Uwe C.},
  title     = {Critical Dynamics: A Field Theory Approach to Equilibrium and Non-Equilibrium Scaling Behavior},
  publisher = {Cambridge University Press},
  address   = {Cambridge},
  year      = {2014},
  doi       = {10.1017/CBO9781139046213},
  isbn      = {978-0-521-84223-5}
}

@article{Gueron1999,
  author  = {Gu{\'e}ron, S. and Deshmukh, M. M. and Myers, E. B. and Ralph, D. C.},
  title   = {Tunneling via Individual Electronic States in Ferromagnetic Nanoparticles},
  journal = {Phys. Rev. Lett.},
  volume  = {83},
  pages   = {4148},
  year    = {1999},
  doi     = {10.1103/PhysRevLett.83.4148}
}

@article{CanaliMacDonald2000,
  author  = {Canali, C. M. and MacDonald, A. H.},
  title   = {Theory of Tunneling Spectroscopy in Ferromagnetic Nanoparticles},
  journal = {Phys. Rev. Lett.},
  volume  = {85},
  pages   = {5623},
  year    = {2000},
  doi     = {10.1103/PhysRevLett.85.5623}
}

@article{Kleff2001,
  author  = {Kleff, S. and von Delft, J. and Deshmukh, M. M. and Ralph, D. C.},
  title   = {Model for Ferromagnetic Nanograins with Discrete Electronic States},
  journal = {Phys. Rev. B},
  volume  = {64},
  pages   = {220401},
  year    = {2001},
  doi     = {10.1103/PhysRevB.64.220401}
}

@article{Lyubshin_2014,
   title={Statistics of spin fluctuations in quantum dots with Ising exchange},
   volume={89},
   ISSN={1550-235X},
   url={http://dx.doi.org/10.1103/PhysRevB.89.201304},
   DOI={10.1103/physrevb.89.201304},
   pages={201304},
   number={20},
   journal={Physical Review B},
   publisher={American Physical Society (APS)},
   author={Lyubshin, D. S. and Sharafutdinov, A. U. and Burmistrov, I. S.},
   year={2014},
   month=May }

@book{risken1996fokker,
  title={The Fokker-Planck Equation: Methods of Solution and Applications},
  author={Risken, Hannes},
  series={Springer Series in Synergetics},
  volume={18},
  edition={2nd},
  year={1996},
  publisher={Springer-Verlag},
  address={Berlin Heidelberg},
  doi={10.1007/978-3-642-61544-3},
  isbn={978-3-540-61530-9}
}

@article{Landauer1975,
   author  = {R. Landauer},
   title   = {Inadequacy of entropy and entropy derivatives in characterizing
             the steady state},
   journal = {Phys. Rev. A}, volume = {12}, pages = {636}, year = {1975}
}

@article{AronChamon2020,
	title = {Landau theory for non-equilibrium steady states},
	pages = {074},
	author = {Aron, Camille and Chamon, Claudio},
	journal = {SciPost Phys.},
	volume = {8},
	year = {2020},
	publisher = {SciPost},
	doi = {10.21468/SciPostPhys.8.5.074},
	url = {https://scipost.org/10.21468/SciPostPhys.8.5.074}
}

@article{Mozyrsky2006,
  author  = {Mozyrsky, D. and Hastings, M. B. and Martin, I.},
  title   = {Intermittent polaron dynamics: {B}orn-{O}ppenheimer approximation out of equilibrium},
  journal = {Phys. Rev. B},
  volume  = {73},
  number  = {3},
  pages   = {035104},
  year    = {2006},
  doi     = {10.1103/PhysRevB.73.035104}
}
\section{End Matter}

\paragraph{Gaussian kernel and open-QCP scaling.}
Expanding the action~\eqref{eqn: bosonic action} to second order in fluctuations about the saddle point gives
\begin{equation}
S^{(2)}=N\!\int\!\frac{\mathrm{d}\omega}{2\pi}
\begin{pmatrix}\delta\phi_{\rm cl}&\delta\phi_q\end{pmatrix}_{-\omega}
\begin{pmatrix}
0&[D^A]^{-1}\\[2pt]
[D^R]^{-1}&[D^{-1}]^K
\end{pmatrix}
\begin{pmatrix}\delta\phi_{\rm cl}\\ \delta\phi_q\end{pmatrix}_{\omega}.
\label{eq:gaussian_action}
\end{equation}
$\bm D$ is fixed by the dot spin polarization matrix
\begin{equation}
\Pi^{\alpha\beta}(t',t)=\frac{i}{2}\,\Tr\!\big[\hat G(t,t')\,\sigma_x\hat\gamma^{\alpha}\,\hat G(t',t)\,\sigma_x\hat\gamma^{\beta}\big],
\label{eq:polarization}
\end{equation}
with $\{\hat{\gamma}^{\text{cl}},\hat{\gamma}^\text{q}\}=\{1,\sigma_{1}\}$ on Keldysh space,
through $[D^{R/A}]^{-1}=\Pi^{R/A}-2$ and $[D^{-1}]^K=\Pi^K$.  In equilibrium, the FDR fixes $\Pi^K=2i\coth(\beta\omega/2)\mathrm{Im}\{\Pi^R\}$. Inserting Eq. ~\eqref{eq:lowfreq_expansion} into Eq.~\eqref{eq:fluct_integral} gives, at $T=V=0$,
\begin{equation}
N\langle\delta\phi^2\rangle\;\sim\;\int_{-\infty}^\infty\!\frac{\mathrm{d}\omega}{2\pi}\,
\frac{\lambda|\omega|}{(a-c\omega^2)^2+\lambda^2\omega^2}
=\frac{1}{2\pi \lambda}\,g\!\Big(\frac{ac}{\lambda^2}\Big),
\label{eq:F_integral}
\end{equation}
omitting the ${\rm cl}$ superscript on $\delta\phi$ and with
\begin{equation}
g(s)=\frac{\pi+2\arctan\!\big(\tfrac{2s-1}{\sqrt{4s-1}}\big)}{\sqrt{4s-1}}
\;\sim\;
\begin{cases}
s^{-1/2}, & s\gg1,\\
-\log s, & s\ll1.
\end{cases}
\label{eq:F_function}
\end{equation}
Integrating out $\delta\phi^q$ in the quadratic action gives the semiclassical equation of motion~\cite{Kamenev2023Jan}
\begin{equation}
2\delta\phi_t-\!\int\! dt'\,\Pi^R_{tt'}\delta\phi_{t'}
=\!\int\! dt'\,\sqrt{-i\Pi^K/2}_{\,tt'}\,\frac{\xi_{t'}}{\sqrt N},
\label{eq:langevin}
\end{equation}
which reduces, with Eq.~\eqref{eq:lowfreq_expansion}, to $c\delta\ddot{\phi}+\lambda\delta\dot\phi+a\delta\phi=\tilde\xi/\sqrt N$. Finally, as $\Gamma\to0$ the Taylor expansion of $a$ in $T$ breaks down. The coefficient $a_1\propto\Gamma$ vanishes, and the thermal correction becomes exponentially suppressed, $a(\Gamma\to0,T,h)\simeq a_0(h)+b\,e^{-h_c/2T}$. This yields an upper crossover temperature $T^\ast\sim-1/\log(\delta h)$ in place of $\sqrt{\delta h/\Gamma}$, producing the reduced fan-like region discussed for the closed LMG model~\cite{hooley2013LMGquasilocalquantum}.

\paragraph{Finite-size scaling exponents at the second-order transition.} The critical order parameter fluctuations,
$\langle\delta\phi^2\rangle\sim N^{\alpha}$, define a finite-size exponent that diagnoses distinct critical regimes. We can extract $\alpha$ from a tree-level scaling analysis of the Keldysh action, in the spirit of Ref.~\cite{DallaTorre2013}. The analysis
reproduces the closed-system QCP value $\alpha=-2/3$ \cite{Dusuel_2005,Dusuel2004}, yields $\alpha=-1$ at the open QCP, and confirms $\alpha=-1/2$ in the thermal/bias-driven regime, agreeing with the finite temperature closed LMG ~\cite{Wilms_2012}. We note that at the critical point, $\langle\delta\phi^2\rangle=\langle\phi^2\rangle=\braket{S_x^2}/N^2$.

The ingredients are the Gaussian kernel of Eq.~\eqref{eq:gaussian_action}, expanded at low frequency
through Eq.~\eqref{eq:lowfreq_expansion}, together with the leading nonlinearity. The lowest-order vertex compatible with the
$\mathbb{Z}_2$ symmetry and with a continuous transition is
$u\,(\phi^{\rm cl})^3\phi^q$, where $u$ is the static quartic coupling 
(fourth derivative of $\mathcal{F}(\phi)$), positive at the second-order transition. Near criticality, the action reads
\begin{align}
S \;=&\; 2N\!\int\! dt\; \phi^q\big(c\,\partial_t^2+\lambda\,\partial_t+a\big)\phi^{\rm cl}
   \;+\; \\&N\!\int\! dt\,dt'\;\phi^q_t\,\Pi^K_{t-t'}\,\phi^q_{t'}
   \nonumber \;+\; N\,u\!\int\! dt\,(\phi^{\rm cl})^3\phi^q .
\label{eq:EM-action}
\end{align}

We rescale $t\to s\,t$, along with $\phi^{cl}\to s^{d_c} \phi^{cl}$, $\phi^q\to s^{d_q}\phi^{q}$, and $N\to s^{d_N}N$. We fix ${d_c,d_q,d_N}$ by demanding that the action remains scale invariant on long time scales. The scaling of equal-time fluctuations follows by matching their scaling to
that of a system of size $s^{d_N} N$ yielding,  $\langle\phi^2\rangle\sim N^{\alpha}$ with $\alpha={2d_c}/{d_N}$.

At criticality, the mass term, $a$, is tuned to zero, and 
scale invariance of the action dictates the following three equations. (i)~\emph{The quartic vertex} requires
$d_N+1+3d_c+d_q=0$.
(ii)~\emph{The kinetic term}. Since $\lambda\simeq 8\Gamma^2/(\pi h^4)$ any $\Gamma>0$ causes the friction, $\lambda \partial_t$, to dominate the
low-frequency retarded inverse fluctuation propagator, leading to $d_N+d_c+d_q=0$. At $\Gamma=0$, however, the friction is absent ($\lambda=0$)
and the inertial term, $c\partial_t^2$ takes over, giving $d_N+d_c+d_q-1=0$.
(iii)~\emph{The relative dimension of the two fields}, fixed by the
FDR $\Pi^K=2F(\omega)\mathrm{Im}\,\Pi^R$.
Comparing the $\phi^q\phi^{cl}$ and $\phi^q\phi^q$ vertices yields
$d_c-d_q=[F]$. At $T=V=0$, $F=\mathrm{sgn}(\omega)$, forcing $d_c=d_q$, whereas, $F(\omega)\sim 2T_\text{eff}/\omega$ at low frequency yielding $d_c= d_q+1$.

Solving Eqs. (i)--(iii) in each regime gives the three different scaling exponents, as shown in the table below.
\begin{table}[h]
\centering
\renewcommand{\arraystretch}{1.25}
\resizebox{\columnwidth}{!}{\begin{tabular}{l c c c c c c}
\hline\hline
Regime & Kinetic & $[F]$ & $d_c$ & $d_q$ & $d_N$ & $\alpha$ \\
\hline
Thermal ($T>0$ or $V>0$)          & $\partial_t$  & $1$ & $-\tfrac12$ & $-\tfrac32$ & $2$ & $-\tfrac12$ \\
Open QCP ($\Gamma>0,\,\{T,V\}=0$)    & $\partial_t$  & $0$ & $-\tfrac12$ & $-\tfrac12$ & $1$ & $-1$ \\
Closed QCP ($\{\Gamma,T,V\}=0$)          & $\partial_t^2$& $0$ & $-1$        & $-1$        & $3$ & $-\tfrac23$ \\
\hline\hline
\end{tabular}
}
\end{table}

Note that in all cases $d_c\geq d_q$, so all other possible $\bm\phi^4$ vertices either scale identically to $(\phi^{cl})^3\phi^q$ (QCP) or become irrelevant when $s\to\infty$ (Thermal).

\begin{figure}[!t]
    \centering
    \includegraphics[width=0.90\linewidth]{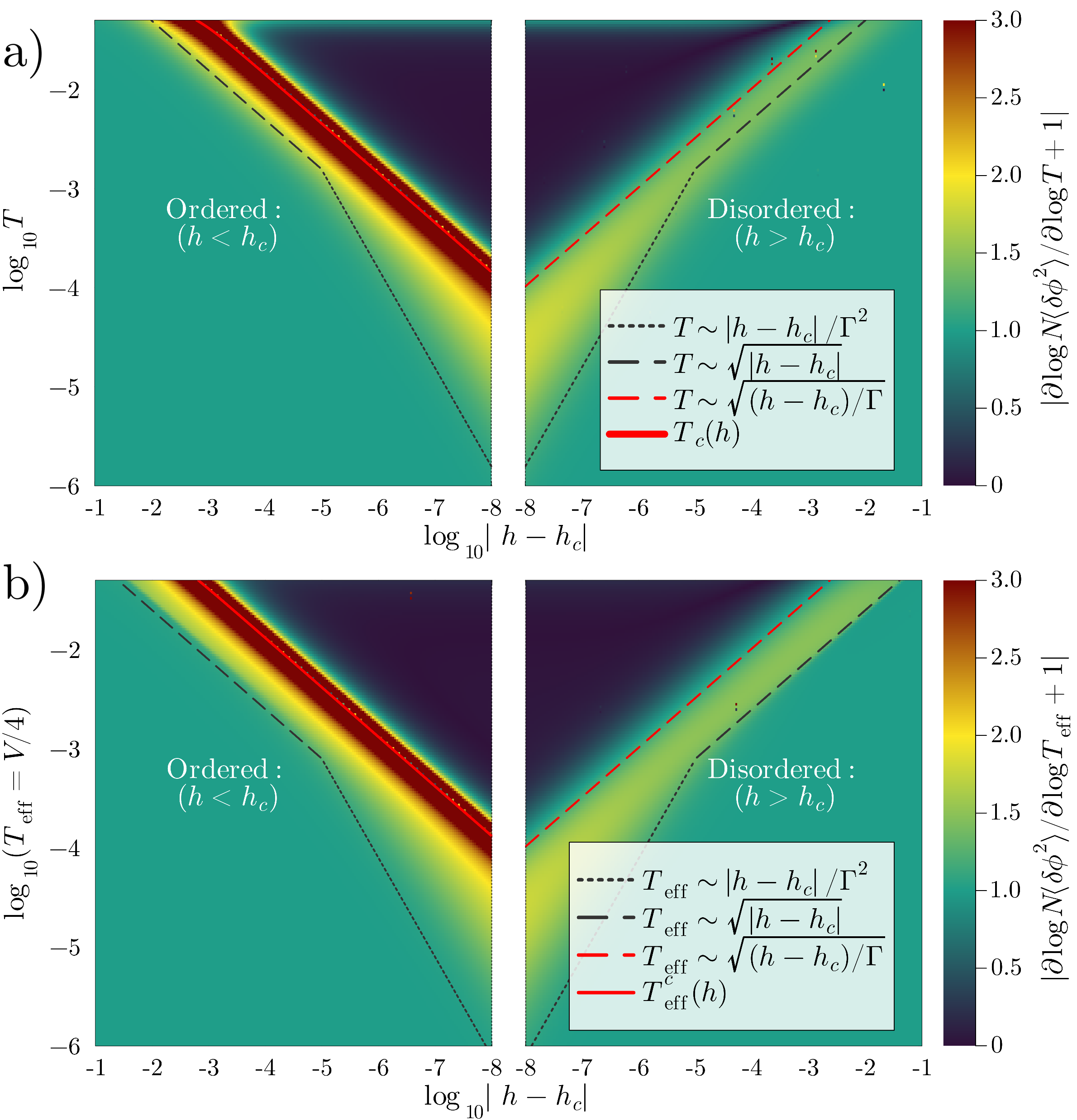}
    \caption[Fluctuation structure near the QCP]{Fluctuation structure near the QCP from numerical integration of Eq.~\eqref{eq:fluct_integral} at $\Gamma=0.05$.
\textbf{(a)} Thermal case ($V=0$): the indicator $|\partial\log N\langle\delta\phi^2\rangle/\partial\log T+1|$ in the $(\,|h-h_c|,T\,)$ plane (log--log), for both $h-h_c>0$ (right) and $h-h_c<0$ (left). The indicator is $0$ in the quantum-critical region [$N\langle\delta\phi^2\rangle\sim1/(\Gamma T)$], $1$ where $N\langle\delta\phi^2\rangle$ is $T$-independent, $2$ in the crossover region ($\sim T$), and diverges on the critical line $T_c(h)$ (red); the color scale is capped at $3$. Overlaid are the crossover curves $T\sim|h-h_c|/\Gamma^2$, $T\sim\sqrt{|h-h_c|}$, and $T\sim\sqrt{|h-h_c|/\Gamma}$ obtained in the main text.
\textbf{(b)} Driven case ($T=0$, varying $V$), plotted against $T_{\rm eff}=V/4$.}
\label{fig:QCP_numerics}
\end{figure}

\paragraph{Numerical determination of the QCP fan.}
The numerical solution of Eq.~\eqref{eq:fluct_integral}, shown in Fig.~\ref{fig:QCP_numerics}(a), agrees with our general arguments for the behavior of $N\braket{\delta\phi^2}$ with $\delta h$ and $T$. The same results extend to the driven case: for $T=0$ [Fig.~\ref{fig:QCP_numerics}(b)], the fluctuation structure is nearly identical to the equilibrium case under the new variable $T_\text{eff}=V/4$ (the $T\to0$, small $V$ limit of Eq.~\eqref{eq:teff_smallV}). In both cases, the quantum critical fan extends to the $\delta h<0$ side, which mirrors the $\delta h>0$ side, with the exception that now the curve separating the ordered and quantum critical regimes is $T_c(h)$ ($V_c(h)$ in the driven case), at which fluctuations diverge.  

\paragraph{Nonequilibrium potential and state selection.}
The time-local description of Eq.~\eqref{eq:langevin_full} treats the order
parameter as slowly varying; similarly to the nonequilibrium Born–Oppenheimer approximation introduced in ~\cite{Mozyrsky2006}. In the dissipative sector this amounts to a
first-order gradient expansion of $\Pi^R$, discarding all higher
time derivatives and, in the noise sector, to discarding the memory of $\Pi^K$. The gradient expansion requires the collective mode to
relax more slowly than the electrons dwell on the dot,
$\gamma_\text{OD}(\phi)\ll\Gamma$. The Markovian approximation
requires the reservoir noise to be classical on the time scale of the mode,
$\gamma_\text{OD}(\phi)\ll T_\text{eff}(\phi)$. The second condition reveals the failure of the description at the equilibrium QCP, where $T_\text{eff}\to0$ and no time-local noise exists ($\Pi^K(\omega)\sim|\omega|$). Within the bistable region both conditions are best satisfied near the
tricritical point (e.g. at $\Gamma=0.35$, chosen for Fig. \ref{fig:new_phase_pdf}), where $\phi$ inherits the critical slowing down of the continuous transition ($\gamma_\text{OD}\to0$).
At the opposite edges of the bistable region
(small $\Gamma$ or $T_\text{eff}$), the two ratios grow and the time-local
description becomes less controlled, so that non-Markovian
 corrections to the non-equilibrium selection criterion are expected.

The stationary solution of the corresponding Fokker--Planck equation is Eq.~\eqref{eq:quasipotential}. At the second-order transition, it gives $\langle\phi^2\rangle\sim N^{-1/2}$, matching our previous scaling analysis. Near the first-order transition, $P_{\rm ss}(\phi)$ develops three peaks [Fig.~\ref{fig:new_phase_pdf}(a)], corresponding to the trivial solution at $\phi=0$ and two symmetric ordered states. The crossover in $\langle|\phi|\rangle=\int d\phi\,|\phi|P_{\rm ss}(\phi)$ sharpens with $N$ [Fig.~\ref{fig:new_phase_pdf}(b)]. The critical point becomes the equal-depth point of $\mathcal W$ [Fig.~\ref{fig:new_phase_1}(a)] rather than of $\mathcal F$; since $T_{\rm eff}$ is largest near $\phi=0$, weight is displaced toward the ordered state. This is an example of Landauer's blowtorch effect from classical non-equilibrium statistical physics \cite{Landauer1975, Landauer1988Oct}.

\paragraph*{Deterministic potential as a sum of single-lead contributions.}
We derive the decomposition of the deterministic potential and identify when it holds, considering a generic Stoner-like dot with  interaction channels $\nu$,
\begin{equation}
H_{\rm dot}=\sum_{i,ss'}\epsilon_{i,ss'}\,d^\dagger_{i,s}d_{i,s'}
-\sum_{\nu}\frac{\gamma_\nu}{2N}\Big(\sum_{i,ss'} d^\dagger_{i,s}\,\sigma^\nu_{ss'}\,d_{i,s'}\Big)^{2},
\label{eq:em_generic_dot}
\end{equation}
where $\sigma^\nu$ selects the channel and $\phi_\nu=\langle\sum_i d^\dagger_i\sigma^\nu d_i\rangle/2N$ is the corresponding order parameter. The model of the main text is the special case of a single channel $\sigma=\sigma_x$ (with $\gamma=J/2$) and a level-independent single-particle term $\epsilon_{i,ss'}=-\tfrac{h}{2}(\sigma_z)_{ss'}$. Decoupling each interaction with a Hubbard--Stratonovich field $\phi_\nu$ and integrating out the fermions, the saddle point $\phi^{\rm cl}_\nu\equiv\phi_\nu$ obeys
\begin{equation}
\begin{aligned}
\phi_\nu=&\frac{1}{2N}\sum_{l}\mathrm{Tr}\!\Bigg\{
\mathbf{\sigma}^\nu\!\!\!
\int\!\!\frac{\mathrm{d}\omega}{2\pi}\,
n_F(\omega-\mu_l)\big[\mathbf{G}^R\bm\Gamma_l\mathbf{G}^A\big]\Bigg\},
\end{aligned}
\label{eq:em_neq_saddle}
\end{equation}
where the trace now runs over both level and spin indices and the dressed propagators are evaluated in the presence of static fields $\sum_\nu2\gamma_\nu\phi_\nu\sigma^\nu$. In general, both $\mathbf{G}^{R/A}$ and $\boldsymbol{\Gamma}_l$ are matrices in the level and spin indices. Note that $\mathbf{G}^{R/A}$ carries the total self-energy of all leads, so the broadening in $\mathbf{G}^{R/A}$ is the total hybridization $\boldsymbol{\Gamma}=\sum_l\boldsymbol{\Gamma}_l$, not the individual $\boldsymbol{\Gamma}_l$ multiplying the Fermi factor.

We compare this with an equilibrium dot coupled to a single reservoir. In the Matsubara formalism, for a dot coupled to one lead at chemical potential $\mu$ and hybridization $\boldsymbol{\Gamma}$, the mean-field free energy is
\begin{equation}
f^{\rm eq}[\boldsymbol{\Gamma},\mu]=\sum_\nu2{\gamma_\nu\phi_\nu^2}
-\frac{1}{\beta N}\sum_{\omega_n}\mathrm{Tr}\ln\!\big[-\mathbf{G}^{-1}(i\omega_n+\mu)\big],
\label{eq:em_feq}
\end{equation}
with $\mathbf{G}^{-1}(i\omega_n)=i\omega_n-\boldsymbol{\epsilon}+2 \sum_\nu\gamma_\nu\phi_\nu\sigma^\nu+\tfrac{i}{2}\mathrm{sgn}(\omega_n)\boldsymbol{\Gamma}$. Its stationary condition $\partial f^{\rm eq}/\partial\phi_\nu=0$ is
\begin{equation}
\phi_\nu-\frac{i}{2N}\,\mathrm{Tr}\!\left\{\sigma^\nu\!\!\int\!\frac{\mathrm{d}\omega}{2\pi}\,
n_F(\omega-\mu)\big[\mathbf{G}^R-\mathbf{G}^A\big]\right\}=0,
\label{eq:em_eq_saddle}
\end{equation}
with the same broadening $\boldsymbol{\Gamma}$ inside $\mathbf{G}^{R/A}$. Note that $\bm G^R-\bm G^A = -i\bm G^R\bm\Gamma\bm G^A$, allowing for a simple comparison with Eq.~\eqref{eq:em_neq_saddle}. If the self-energies share a common structure, $\boldsymbol{\Gamma}_l=u_l\,\boldsymbol{\Gamma}$ (with $\sum_l u_l=1$), the driven conditions become a weighted average of equilibrium ones and admit a common potential:
\begin{equation}
\frac{\partial\mathcal F}{\partial\phi_\nu}=0,
\qquad
\mathcal F=\sum_lu_l\,f^{\rm eq}[\boldsymbol{\Gamma},\mu_l],
\label{eq:em_decomposition}
\end{equation}
for all channels simultaneously. In the wide-band limit the common structure condition is trivially satisfied, since $\boldsymbol{\Gamma}_l$ is a scalar constant.  It generically fails in spatially extended systems, where different leads couple with distinct spatial dependence and the potential therefore no longer simplifies to a weighted average of equilibrium free energies.

\end{document}